\documentclass[twocolumn,showpacs,preprintnumbers,amsmath,amssymb,superscriptaddress, array]{revtex4}

\usepackage{graphics}
\usepackage{dcolumn}
\usepackage{bm}

\newcommand{\const}{{\rm\, const}}

\input epsf
\usepackage{amsmath,amssymb,epsfig,placeins,subfigure,wrapfig,indentfirst}

\begin{document}

\title{Passage of radiation through wormholes}

\author{Andrey Doroshkevich}
\affiliation{Astro Space Center, Lebedev Physical
  Institute, Russian Academy of Sciences, Moscow, Russia}
\author{Jakob Hansen}
\affiliation{Waseda University, Department of Physics, Tokyo, Japan}
\author{Igor Novikov}
\affiliation{Astro Space Center, Lebedev Physical
  Institute, Russian Academy of Sciences, Moscow, Russia}
\affiliation{Niels Bohr Institute, Copenhagen,
  Denmark}
\affiliation{Kurchatov Institute,  Moscow, Russia}
\author{Alexander Shatskiy}
\affiliation{Astro Space Center, Lebedev Physical
  Institute, Russian Academy of Sciences, Moscow, Russia}

\date{\today}

\begin{abstract}
We investigate numerically the process of the passage of a
radiation pulse through a wormhole and the subsequent evolution of
the wormhole that is caused by the gravitational action of this
pulse. The initial static wormhole is modeled by the spherically
symmetrical Armendariz-Picon solution with zero mass. The
radiation pulses are modeled by spherically symmetrical shells of
self-gravitating massless scalar fields. We demonstrate that the
compact signal propagates through the wormhole and investigate the
dynamics of the fields in this process for both cases: collapse of
the wormhole into the black hole and for the expanding wormhole.
\end{abstract}


\maketitle

\section{Introduction}
\label{sec:0}

One of the most interesting features of the theory of general
relativity is the possible existence of spacetimes with wormholes
\cite{Wheeler55,Wheeler57,Misner57,dop-0,Visser95,Ellis73}. The
wormholes are topological tunnels which connect different
asymptotically flat regions of our Universe or different universes
in the model of Multiverse \cite{Carr07}.

Recently the hypothesis that some known astrophysical objects
(e.g. the active nuclei of some galaxies) could be entrances to
wormholes was considered by Kardashev, Novikov and Shatskiy
\cite{Kardashev06,Kardashev07} and by Shatskiy \cite{Shatskii07}.
If some of these wormholes are traversable, then radiation and
information, at least in principle, can pass between two regions
of the Universe or from one universe to another universe in the
model of the Multiverse. Furthermore, if a wormhole exists, in
principle it is possible to transform it into a time machine
(space-time with closed time-like curves) \cite{Morris88,
Novikov89}. Therefore, it is very important for both the theory
and the development of the hypothesis about the real existence of
the wormholes to analyze the physics of the passage of radiation
through a wormhole. According to the wormhole models of general
relativity, in the absence of any special matter, the throat of
the wormhole pinches off so quickly that it cannot be traversed
even by a test signal moving with the velocity of
light~\cite{Frolov98}. In order to prevent the shrinking of a
wormhole and to make it traversable, it is necessary to thread its
throat with so-called "exotic matter", which is matter that
violates the averaged null energy condition (see
\cite{Visser95,Thorne93,Flanagan96}). Many questions associated
with the existence of wormholes and with exotic matter remain
unsolved, e.g. its stability against various processes, and
different views have been expressed in the literature (see for
example \cite{Visser95, Frolov98, Bronnikov06, Lemos-Lobo}). We
will not discuss this here. Different types of wormholes may exist
depending on the type of exotic matter in their throats (see
\cite{Armendariz-Picon02, Kardashev07,Shatskiy08}). For example,
it could be a "magnetic exotic matter"\, in which the main
component is a strong, ordered, magnetic field plus a small amount
of a "true exotic matter"\, (see \cite{Kardashev07}). Another type
of exotic matter is a "scalar exotic matter"\, in the form of a
scalar field with a negative energy density (see
\cite{Armendariz-Picon02}). Finally, it can be in the form of a
mixed "magnetic-negative dust exotic matter"\, type in which it is
a mixture of an ordered magnetic field and dust (matter with zero
pressure) with negative matter density (see \cite{Shatskiy08}).
The physical properties of different types of wormholes are
different (see for example \cite{Shatskiy08}). The possible
consequences of the passage of radiation through a wormhole with a
"scalar exotic matter"\, was considered, using an analytical
approach, by Doroshkevich, Kardashev, D. Novikov and I. Novikov in
\cite{Doroshkevich08}, where a toy model was used to accept some
very artificial hypotheses about the final state of the wormhole.
The results obtained in \cite{Doroshkevich08} allowed the authors
to reach some important conclusions, but of course without
numerical computations, the real dynamics of the processes of
radiation propagation through a wormhole could not be considered.
Other important analytical analyses of the existence and evolution
of the wormholes and their possible transformation into black
holes (BH) was seen in
\cite{dop-1,dop-2,dop-3,dop-4,dop-5,Shatskiy07,Dokuchaev04,Dokuchaev05}.
Other important results are also seen in the papers
\cite{dop-9,dop-10,dop-11}.

The new period of the investigation of the wormholes began from
the seminal papers \cite{dop-6,dop-7}, where the wormhole dynamics
were analyzed using numerical simulations. Paper \cite{dop-6} is
devoted mainly to the evolution of an initial static wormhole
maintained by the scalar exotic matter under the gravitational
action of the scalar radiation pulse passing the wormhole from one
asymptotically flat region to other one.  In paper \cite{dop-7}
the authors analyze the nonlinear evolution of the wormhole
perturbed by the scalar field.

The goal of this paper is to continue the line of the
investigation of these two works. We focus on the dynamics of the
scalar fields during the nonlinear stages of the wormhole evolution
under the gravitational action of the ingoing pulse of the scalar
radiation (both exotic and usual). These dynamics can explain some
results of the papers \cite{dop-6,dop-7}.

This paper is organized as follows:

In section~\ref{sec:I} we present the initial model of the
wormhole. In section~\ref{sec:II} the field equations are written
out. The initial value problem is discussed in
section~\ref{sec:III}. In section~\ref{sec:IV} we present the
numerical solution of the field equation which describe the
passage of the signal through the wormhole and its subsequent
evolution with the formation of a BH. In section~\ref{sec:V} we
analyze the case of a strong signal. The passage of the signal of
the exotic scalar field is considered in section~\ref{sec:VI}.
Finally we summarize our conclusions in section~\ref{Conclusion}.
Mathematical details and the details of the numerical code are
given in appendices~\ref{sec:app0}-\ref{sec:appB}.

\section{The Model}
\label{sec:I}

We study, numerically, the evolution of a simple spherical model
of a wormhole maintained by an exotic scalar field and perturbed
by an in-falling massless, self-gravitating scalar field. This
initially in-falling scalar field is a pulse in the form of a
spherical layer with some width and amplitude. It imitates a pulse
of radiation directed into the wormhole and propagating with
velocity of light. We do not impose that the in-falling scalar
field is weak. We will consider two cases: when this in-falling
field has a positive energy density (imitating real radiation),
and when it has a negative energy density (imitating exotic
radiation). As an initial static wormhole, let us consider the
special case of the Armendariz-Picon solution of the Einstein
equations for the spherical static state with the effective mass
equal to zero \cite{Armendariz-Picon02,Ellis73}. Another name of
this wormhole is Moris-Thorne's (MT) wormhole. The corresponding
line element for such a wormhole can be written as follows:

\begin{equation}
  \label{eq:lineelement}
ds^{2^{(MT)}}= - d\tau^2 +dR^2 + r^2 d\Omega^2,\,\,\, r^{2^{(MT)}}
= Q^2 + R^2,
\end{equation}
where $R$ is running from $-\infty$ to $+\infty$, $Q$ is a
characteristic of the strength of the exotic supporting field
$\Psi$ and $d\Omega^2=d\theta^2+\sin^2\theta\, d\varphi^2$ is the
line element on the unit two-sphere. The general equations will
be given below. Here we give the non-zero components of the
energy-momentum tensor of the $\Psi$-field for the solution
\eqref{eq:lineelement}:
\begin{subequations}
\label{eq:2}
\begin{eqnarray}
T^{\tau^{(MT)}}_{\tau} &=& \frac{Q^2}{8\pi r^4}\\
T^{R^{(MT)}}_{R} &=& -\frac{Q^2}{8\pi r^4} \\
T^{\theta^{(MT)}}_{\theta} &=& T^{\varphi^{(MT)}}_{\varphi} =
\frac{Q^2}{8\pi r^4}
\end{eqnarray}
\end{subequations}
The narrowest part of the wormhole \eqref{eq:lineelement} is at
$R=0$. This throat corresponds to $r=Q$. The physical analysis of
the metric \eqref{eq:lineelement} is given in
\cite{Doroshkevich08}. Here we emphasize that for the metric
\eqref{eq:lineelement}, in this reference frame, there is no
gravitational acceleration at any point in the three-dimensional
space, but the wormhole (two asymptotically flat three-dimensional
regions joined by a three-dimensional tunnel) nevertheless exists
due to the distribution \eqref{eq:2} of the exotic scalar field.
Of course there are not any apparent and event horizons in such an
object and test signals can pass through the tunnel in both
directions.

The initial motivations of our choice of the MT wormhole as an
initial static wormhole were the following.

(1) The corresponding solutions
(\ref{eq:lineelement})-(\ref{eq:2}) of the Einstein equations are
very simple and elegant.

(2) It was declared in~\cite{Armendariz-Picon02} that the MT
wormhole is stable.

However the analytical \cite{dop-8} and numerical \cite{dop-7}
analysis (see also Appendix~\ref{sec:app0}) demonstrates that the
solutions (\ref{eq:lineelement})-(\ref{eq:2}) are unstable against
small spherical perturbations. Our numerical experiments confirms
this conclusion.

In the light of this fact, the analysis of the possibility of the
passage of the signals through the MT wormhole and their
distortions have a special interest.

\section{Field Equations}
\label{sec:II}

For the numerical analysis we will use the double null
coordinates. The general line element in these coordinates can be
written as:
\begin{equation}
  \label{eq:3}
ds^2 = -2 e^{2 \sigma (u,v)} du\, dv + r^2 (u,v) d\Omega^2,
\end{equation}
where $\sigma (u,v)$ and $r(u,v)$ are functions of the null
coordinates $u$ and $v$ (in- and out-going respectively). The
non-zero components of the Einstein tensor are:
\begin{subequations}
\label{eq:4}
\begin{eqnarray}
 G_{uu} &=& \frac{4 r_{,u} \sigma_{,u} -2r_{,uu} } {r}\\
 G_{vv} &=& \frac{4 r_{,v} \sigma_{,v} -2r_{,vv} } {r}\\
 G_{uv} &=& \frac{e^{2\sigma} + 2r_{,u} r_{,v} + 2 r r_{,uv}}{r^2}\\
 G_{\theta\theta} &=& -2 e^{-2\sigma} r (r_{,uv} + r \sigma_{,uv})\\
 G_{\varphi\varphi} &=& -2 e^{-2\sigma} r \sin^2\theta (r_{,uv} + r
\sigma_{,uv})
\end{eqnarray}
\end{subequations}

The energy-momentum tensor can be written as a sum of
contributions from the exotic scalar field $\Psi$ and the ordinary
scalar field $\Phi$ with the positive energy density: $T_{\mu\nu}
= T_{\mu\nu}^{\Psi} + T_{\mu\nu}^{\Phi}$ :

\begin{equation}
\label{eq:5}
   T_{\mu\nu}^{\Psi} =\frac{-1}{4\pi} \begin{pmatrix}
   \Psi^2_{,u} & 0 & 0 & 0 \\
   0 & \Psi^2_{,v} & 0 & 0 \\
   0 & 0 & r^2 e^{-2\sigma}\Psi_{,u}\Psi_{,v} & 0 \\
   0 & 0 & 0 & r^2 \sin^2\theta\, e^{-2\sigma}\Psi_{,u}\Psi_{,v}
   \end{pmatrix}
   \end{equation}

\begin{equation}
\label{eq:6}
   T_{\mu\nu}^{\Phi} =\frac{+1}{4\pi} \begin{pmatrix}
   \Phi^2_{,u} & 0 & 0 & 0 \\
   0 & \Phi^2_{,v} & 0 & 0 \\
   0 & 0 & r^2 e^{-2\sigma}\Phi_{,u}\Phi_{,v} & 0 \\
   0 & 0 & 0 & r^2 \sin^2\theta\, e^{-2\sigma}\Phi_{,u}\Phi_{,v}
   \end{pmatrix}
   \end{equation}
The $u-u$. $v-v$, $u-v$ and $\theta-\theta$ components of the
Einstein equations respectively are (with $c=1$, $G=1$):

\begin{eqnarray}
& &  r_{,uu} - 2\, r_{,u}\,\sigma_{,u} - r\, \left(\Psi_{,u} \right)^2 + r\,
\left(\Phi_{,u} \right)^2 =0  \label{eq:7}\\
& &  r_{,vv} - 2\, r_{,v}\,\sigma_{,v} - r\, \left(\Psi_{,v} \right)^2 + r\,
\left(\Phi_{,v} \right)^2 =0   \label{eq:8}\\
& &  r_{,uv} +\frac{r_{,u} r_{,v}}{r}+\frac{e^{2\sigma}}{2r} = 0  \label{eq:9}\\
& &  \sigma_{,uv} - \frac{r_{,v} r_{,u}}{r^2} -
\frac{e^{2\sigma}}{2r^2} - \Psi_{,u}\Psi_{,v}  +
\Phi_{,u}\Phi_{,v}  = 0   \label{eq:10}
\end{eqnarray}

The scalar fields satisfy the Gordon-Klein equation
$\nabla^{\mu}\nabla_{\mu}\Psi = 0$ and
$\nabla^{\mu}\nabla_{\mu}\Phi = 0$, which in the metric
\eqref{eq:3} become:
\begin{eqnarray}
& & \Psi_{,uv} + \frac{1}{r} \left( r_{,v}\Psi_{,u} + r_{,u}\Psi_{,v} \right) =
0 \label{eq:11a}\\
& & \Phi_{,uv} + \frac{1}{r} \left( r_{,v}\Phi_{,u} +
r_{,u}\Phi_{,v} \right) = 0 \label{eq:11b}
\end{eqnarray}

Equations \eqref{eq:9}-\eqref{eq:11b} are evolution equations
which are supplemented by the two constraint equations
\eqref{eq:7} and \eqref{eq:8}. It is noted that none of these
equations depends directly on the scalar fields $\Psi$ and $\Phi$,
but only on their derivatives, i.e. the derivative of the scalar
field is a physical quantity, while the absolute value of the
scalar field itself is not. Specifically we note the
$T_{uu}^{\Psi} = - (\Psi_{,u})^2 / (4\pi)$  and $T_{vv}^{\Psi} = -
(\Psi_{,v})^2 / (4\pi)$ components of the energy-momentum tensor
\eqref{eq:5} and $T_{uu}^{\Phi} = (\Phi_{,u})^2 / (4\pi)$  and
$T_{vv}^{\Phi} = (\Phi_{,v})^2 / (4\pi)$ components of the
energy-momentum tensor \eqref{eq:6}, which are part of the
constraint equations. Physically $T_{uu}$ and $T_{vv}$ represents
the flux of the scalar field through a surface of constant $v$ and
$u$ respectively. These fluxes will play an important role in our
interpretation of the numerical results in sections~\ref{sec:IV},
\ref{sec:V} and~\ref{sec:VI}.

\section{Initial value problem}
\label{sec:III}

We wish to numerically evolve the unknown functions $r(u,v)$,
$\sigma (u,v)$, $\Phi (u,v)$ and $\Psi (u,v)$ throughout some
computational domain. We do this by following the approach of
\cite{Burko97c, Burko02b, Burko97, Burko99-1, Burko98-1,
Burko99-2, Burko02-1, Hansen05, Pretorius04} to numerically
integrate the four evolution equations
\eqref{eq:9}-\eqref{eq:11b}. These equations form a well-posed
initial value problem in which we can specify initial values of
the unknowns on two initial null segments, namely an ingoing
(${v=v_0 = constant}$) and an outgoing (${u=u_0 = constant}$)
segment. We impose the constraint equations \eqref{eq:7} and
\eqref{eq:8} on the initial segments. Consistency of the evolving
fields with the constraint equations is then ensured via the
contracted Bianchi identities \cite{Burko97}, but we use the
constraint equations throughout the domain of integration to check
the accuracy of the numerical simulation.

Our choice of the initial values corresponds to the following
physical situation; There is an MT-wormhole and at some distance
from one of the entrances, at the initial moment there is a rather
narrow spherical layer of an in-falling scalar field. There is not
any radiation coming to the wormhole from the side of the other
entrance. As mentioned, we specify the initial values on two
initial segments, namely $u=constant$ and $v=constant$. In the
next section \ref{sec:IV} we consider the case where the
in-falling layer consists of the $\Phi$-field. Later in the
section~\ref{sec:VI} we consider the case of the in-falling layer
consisting of the $\Psi$-field.

For the case
of the layer from the $\Phi$-field, the initial condition could be
specified as follows:

First, let us consider the static MT solution
\cite{Armendariz-Picon02}:
\begin{equation}
\label{eq:MTsolution0} \Psi_{,R}^{(MT)} = \frac{Q}{r^2} =
\frac{Q}{Q^2 + R^2} \Rightarrow \Psi^{(MT)} = \arctan \left(
\frac{R}{Q} \right)
\end{equation}

In $u-v$ coordinates we can write the MT solution as:
\begin{eqnarray}
& & r(u,v) = \sqrt{Q^2 + \frac{1}{4}\left( v-u \right)^2}
\label{eq:MTsolution1}
\\
& & e^{-2\sigma^{(MT)}} = 2 \label{eq:MTsolution2} \\
& & \Psi^{(MT)} = \arctan \left( \frac{v-u}{2 Q} \right)
\label{eq:MTsolution3}
\\
& & \Phi^{(MT)} = 0
\end{eqnarray}

Equations \eqref{eq:MTsolution1} - \eqref{eq:MTsolution3}
completely specifics the MT-wormhole in $u-v$ coordinates in all
of our computational domain including the initial surfaces
$u=constant$ and $v=constant$.

Now, let us consider the more general case with non-trivial $\Phi$
and $\Psi$ scalar fields. In sections \ref{sec:IV} and \ref{sec:V}
we consider the case of a non-zero $\Phi$ field and in section
\ref{sec:VI} we consider the case of a perturbed MT $\Psi$ field
on the initial outgoing $u=constant$ surface. In general we are
free to choose the $\Phi$ and $\Psi$ fields (and their
derivatives) on the initial surfaces in any way we wish. Our
choices for the $\Phi$ and $\Psi$-fields are described in sections
\ref{sec:IV} - \ref{sec:VI}. We are also free to choose $r(u,v)$
on the initial surfaces, this merely expresses the gauge freedom
associated with the transformation $u\rightarrow \tilde{u}(u),
v\rightarrow \tilde{v}(v)$ (the line element \eqref{eq:3} and the
equations \eqref{eq:7}-\eqref{eq:11b} are invariant to such a
transformation). We continue to use eq. \eqref{eq:MTsolution1} on
the initial surface as our choice of gauge. Hence, the only
variable left for us to specify on the initial surfaces is
$\sigma$. This can easily be found by integrating the constraint
equations eq. \eqref{eq:7} and \eqref{eq:8}, which ensures that
the constraint equations are satisfied on the initial
hypersurfaces. Specifically on the outgoing $u=u_0$  hypersurface
it is found by the integral:
\begin{equation}
\label{eq:sigma1} \sigma(u_0,v) =
\ln\left(\frac{1}{\sqrt{2}}\right) + \int\limits_0^{v}
\frac{r_{,vv} - r\, \left(\Psi_{,v} \right)^2 + r\,
\left(\Phi_{,v} \right)^2 }{ 2\, r_{,v}}dv
\end{equation}

The initial conditions on the initial ingoing hypersurface is set
equal to the MT solution for all simulations. Hence, by specifying
a distribution of the scalar fields $\Phi$ and $\Psi$ on the
initial null segments, choosing a gauge and charge parameter
"$Q$"\, we can specify complete initial conditions on the initial
null segments. Using a numerical code (described in appendix
\ref{sec:appA}), we can then use the evolution equations, eqs.
\eqref{eq:9}-\eqref{eq:11b} to evolve the unknown functions
throughout the computational domain.

\section{Physics of the passage of the $\Phi$-field pulse through the MT-
wormhole}
\label{sec:IV}

We investigate the full nonlinear processes arising in the case of
propagation of a compact pulse of the scalar field through an
MT-wormhole, using results of our numerical simulations. Our
numerical code is described in appendix~\ref{sec:appA} and tested
in appendix~\ref{sec:appB}. In this section we analyze the passage
of the $\Phi$-field pulse. In section \ref{sec:VI} we investigate
the passage of the $\Psi$-field pulse. Throughout this paper, the
constant $Q$ (see eq. \eqref{eq:lineelement}), prior to influence
from scalar pulses, has initial value of $Q=1$. The computational
domain for all results throughout this paper is $v=[8,28]$ and
$u=[0,20]$. The
flux of the scalar $\Phi$-field into the wormhole is specified
along initial ${u = u_0 = 0}$ outside the wormhole in the
following way:
\begin{equation}
\label{eq:12} \Phi_{,v} (u_0 ,v) = A^{\Phi} \sin^2 \left( \pi
\frac{v-v_1}{v_2 - v_1}\right)
\end{equation}
where $v_1$ and $v_2$ marks the beginning and end of the ingoing
scalar pulse, respectively, and $A^{\Phi}$ measures the amplitude of the
pulse. Before and after the pulse, at ${v< v_1}$ and ${v > v_2}$
the flux through ${u=u_0}$ is set equal to zero, i.e. ${\Phi_{,v}
(u_0 , v )= 0}$. The flux of the scalar $\Phi$-field through
initial segment ${v=v_0}$ is set equal to zero: ${\Phi_{,u} (u,v_0
) = 0}$.
The expression \eqref{eq:12} can readily
be integrated to give:
\begin{equation}
\label{eq:13} \Phi (u_0, v) = \frac{A^\Phi}{4\pi} \left(
2\pi\left( v-v_1\right) - (v_2 - v_1 )
\sin\left(2\pi\frac{v-v_1}{v_2 - v_1} \right)   \right)
\end{equation}

Note that we formulate the initial condition directly for the flux
${T_{vv}=(\Phi_{,v})^2/(4\pi)}$ of the scalar field through the
surface ${u=u_0}$, rather than for $\Phi$ itself since $T_{vv}$
has the direct physical meaning. Also remember from section
\ref{sec:III} that once the flux through the two initial surfaces
has been chosen, all other initial conditions are determined by
the model. In the examples of the results of our computations we
specify ${v_1 =9}$ and ${v_2 = 11}$. In our
computations we vary the amplitude $A^{\Phi}$ of the pulse.

\begin{figure}
\includegraphics[width=0.49\textwidth]{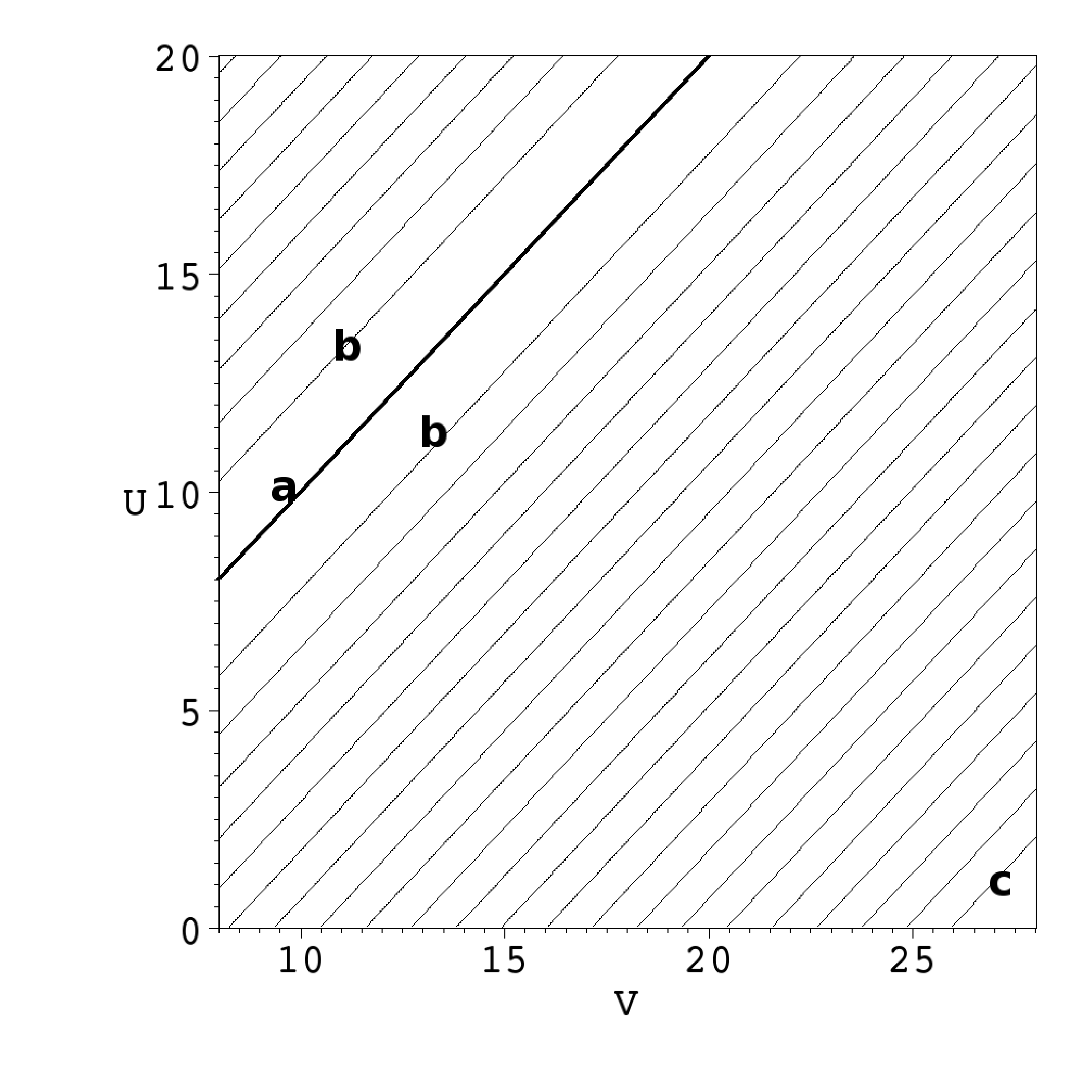}
\caption{Static MT-wormhole. Plot shows contour lines of constant
$r$. Thick diagonal line "a"\, marks the throat of the wormhole
(corresponding to the diagonal $u=v$) with $r=1$. The thin lines
marks lines of constant $r$ increasing outwards from the throat of
the wormhole with a constant spacing between lines of $\Delta r =
0.5$ (i.e. line "b"\, marks $r=1.5$ up to line "c"\, at $r=12$.).
\label{fig:1}}
\end{figure}

In Fig.~\ref{fig:1} is seen the static case of the MT metric
\eqref{eq:lineelement} in $u-v$ coordinates, without any
perturbations of the scalar fields. The throat ${r=1}$ corresponds
to the diagonal ${u=v}$.

\begin{figure*}
\subfigure[Lines of constant $r$, ranging from $r=0$ to $r=5$ with
a spacing between lines of $\Delta r = 0.25$ (line "a"\, marks
$r=1.25$, line "b"\, marks $r=5$ and line "c"\, marks $r=0.75$).
The thick dotted lines marks the position of the apparent
horizons. The thick fully drawn line marks the position of the
singularity $r=0$, hence the region marked "X"\, is outside of the
computational domain.
\label{fig:2a}]{\includegraphics[width=0.49\textwidth]{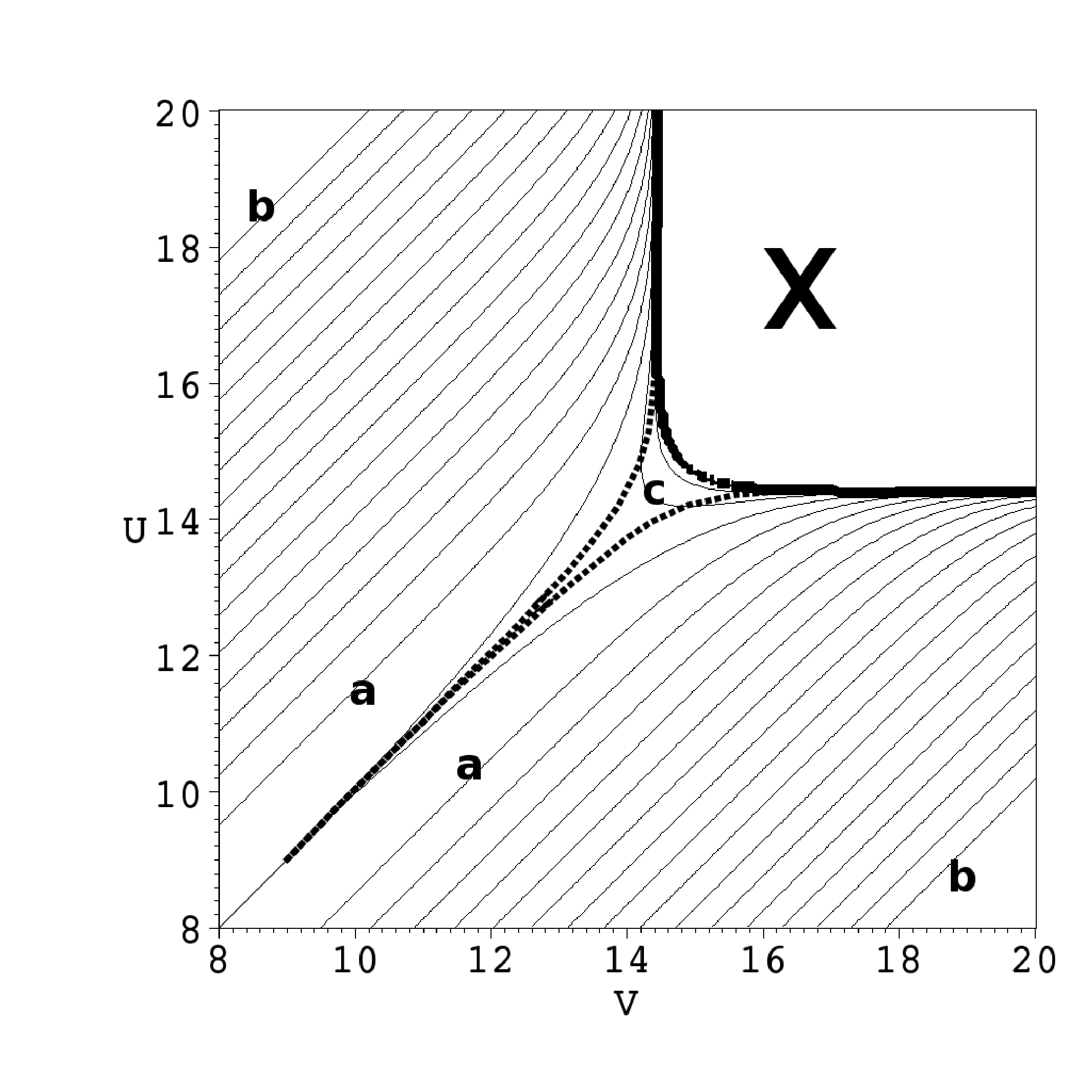}}
\subfigure[Plot of the metric coefficient $e^{2\sigma (L)}$ as a
function of distance parameter $L=\sqrt{(v-v_0)^2 + (u-u_0)^2}$
measuring the distance along the wormhole throat at $u=v$ (with
$L=0$ corresponding to where the $\Phi$-pulse first reaches the
wormhole throat, i.e. $v_0=u_0=9$).
 \label{fig:2b}]{\includegraphics[width=0.49\textwidth]{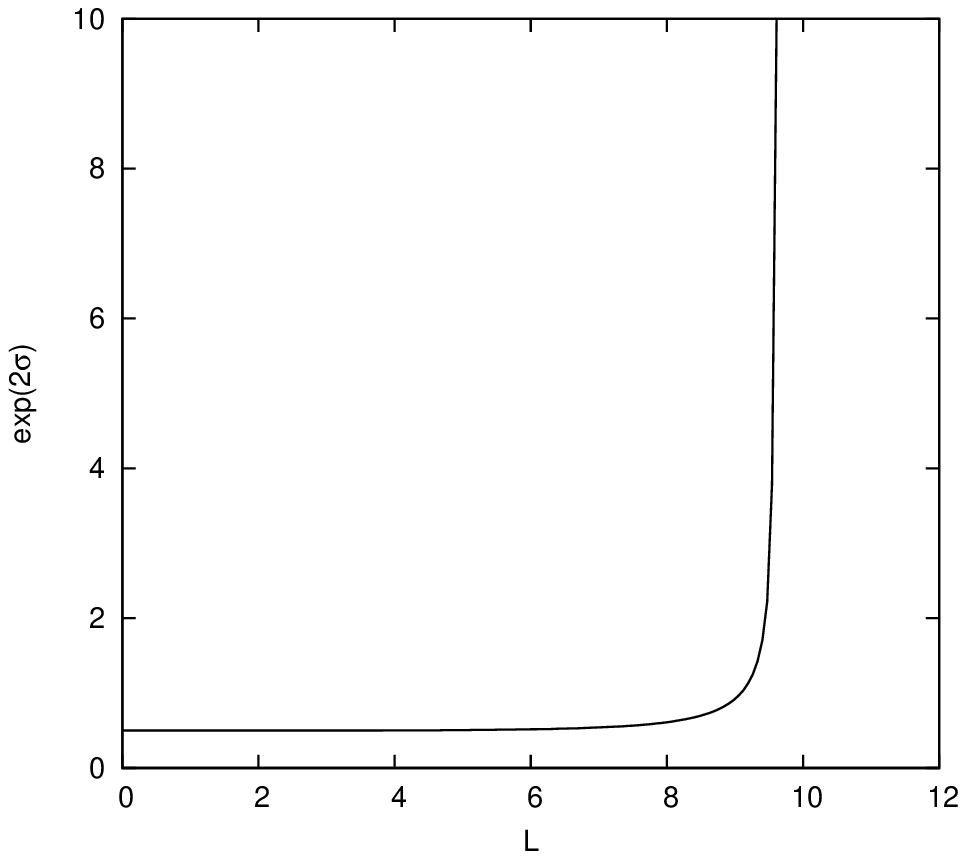}}
\caption{{\label{fig:2} Results of simulation with a non-zero
$\Phi$-field modeled after eq. \eqref{eq:13} with $v_1 = 9, v_2 =
11$ and amplitude $A^{\Phi} = 0.01$.}}
\end{figure*}
\begin{figure*}
\subfigure[Contour lines of the ingoing $\Phi$ flux,
$T_{vv}^{\Phi}=const.$. The lines "a", "b"\, and "c"\, denotes
regions of $T_{vv}^{\Phi}=10^{-3}, T_{vv}^{\Phi}=10^{-4}$ and
$T_{vv}^{\Phi}=3\cdot 10^{-6}$ respectively.
\label{fig:3a}]{\includegraphics[width=0.49\textwidth]{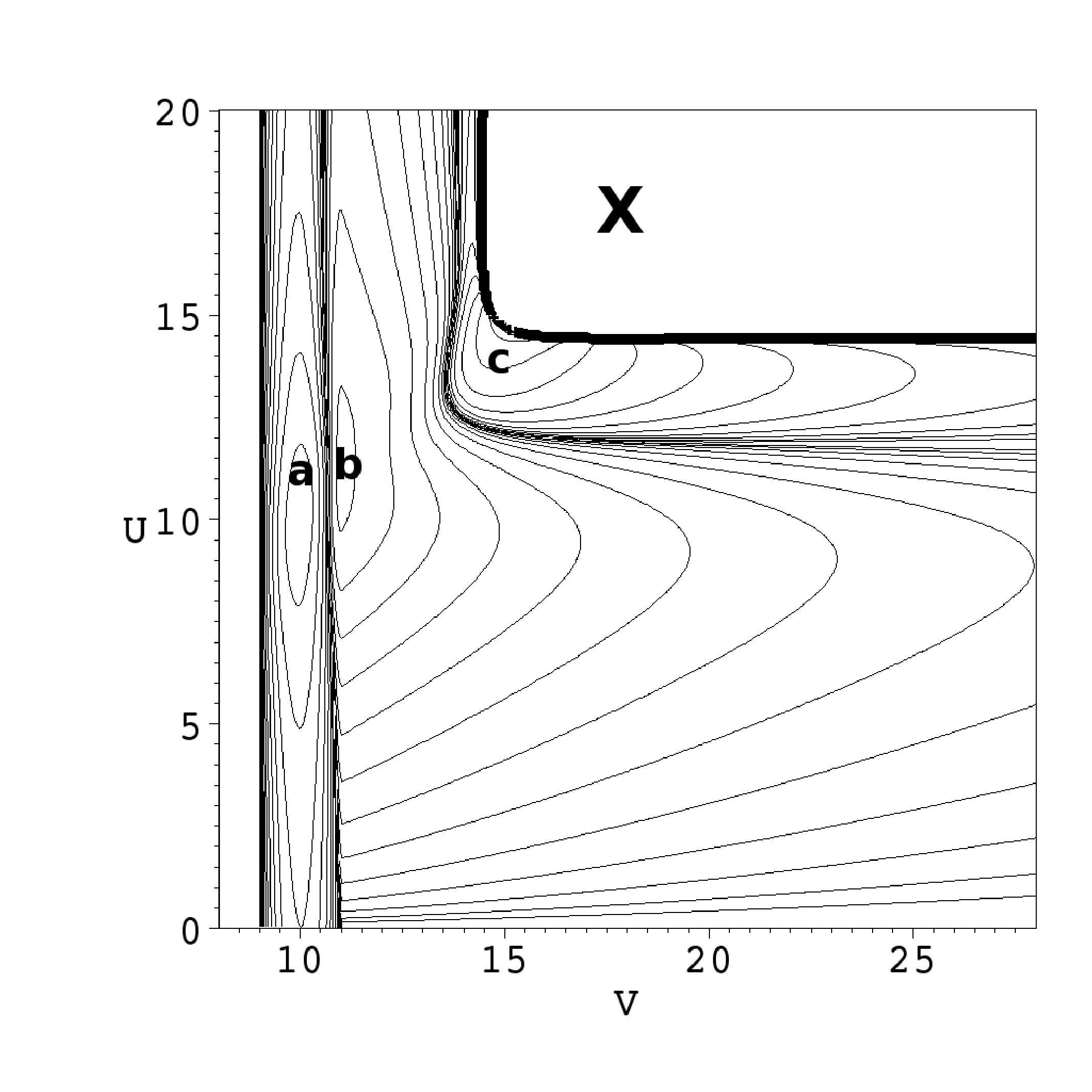}}
\subfigure[Contour lines of the outgoing $\Phi$ flux,
$T_{uu}^{\Phi}=const.$. The lines "a", "b"\, and "c"\, denotes
regions of $T_{uu}^{\Phi}=3\cdot 10^{-5}, T_{uu}^{\Phi}=3\cdot
10^{-5}$ and $T_{uu}^{\Phi}=10^{-9}$ respectively. The letter "d"
identifies a region where the outgoing flux increases towards the
$r=0$ singularity, the contour line closest to the $r=0$
singularity corresponds to $T_{uu}^{\Phi}=10^{-4}$.
\label{fig:3b}]{\includegraphics[width=0.49\textwidth]{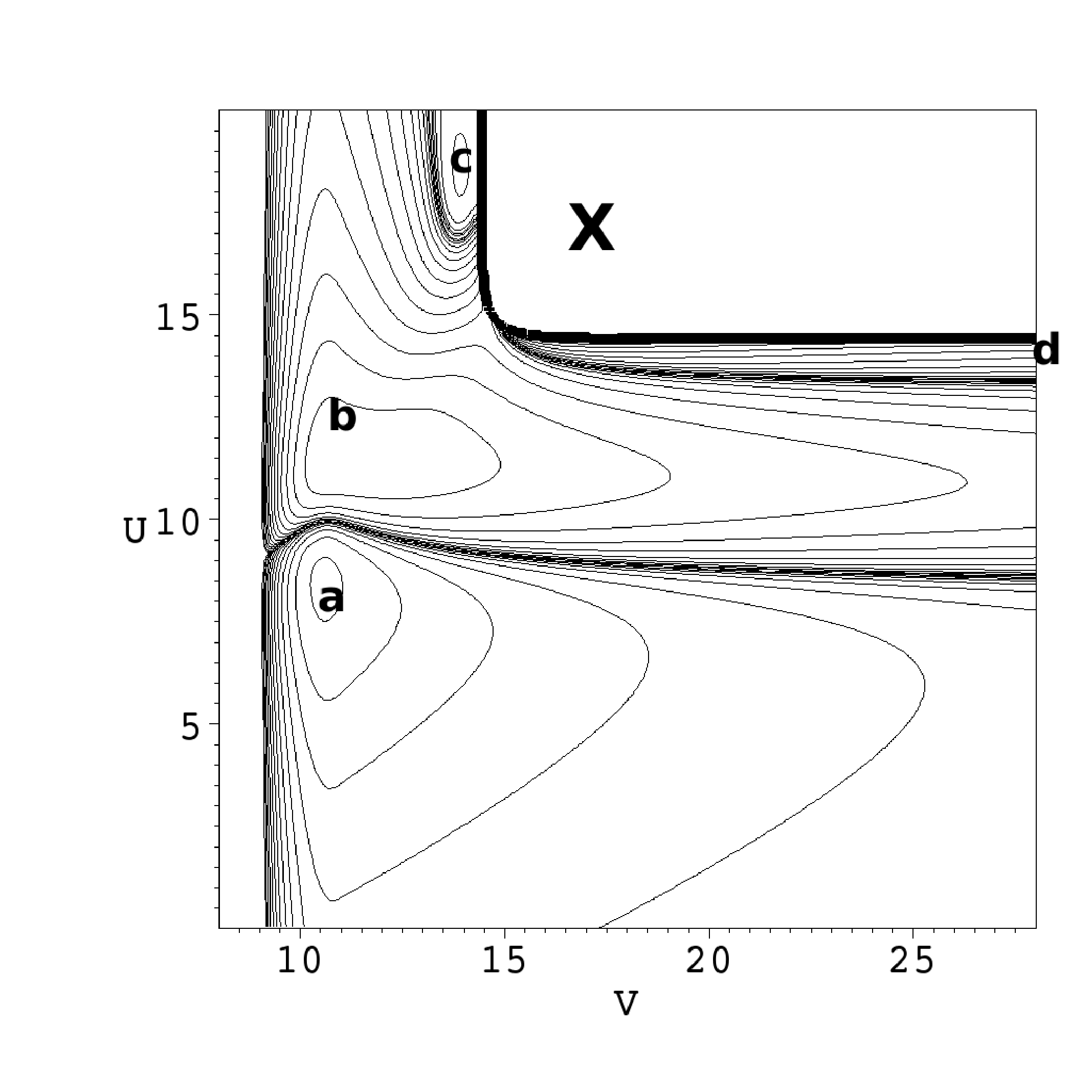}}
\caption{{\label{fig:3} Contour lines of the in- and out-going
$\Phi$ fluxes for the simulation with $v_1 = 9, v_2 = 11$ and
amplitude $A^{\Phi} = 0.01$. In both plots lines are
logaritmically spaced with a factor $10^{\frac{1}{2}}$ between the
lines. The thick solid line marks the location of the $r=0$
singularity and the region "X"\, marks the area outside of the
computational domain. }}
\end{figure*}

In Fig.~\ref{fig:2} one can see a typical example of the evolution
of the wormhole due to the action of the in-falling $\Phi$ scalar
field. Initial values for this $\Phi$-field are ${v_1 = 9}$, ${v_2
= 11}$, ${A^{\Phi}=0.01}$. Fig.~\ref{fig:2a} shows the evolution
of the lines ${r=\const}$. As is seen in Fig.~\ref{fig:2a}, the
evolution corresponds to contraction of the throat down to
${r=0}$, and contraction of the whole wormhole. The real
singularity of the space-time arises at ${r=0}$. It is seen from
the fact that when we come to ${r=0}$, the metric coefficient
$e^{2\sigma}$ in formula \eqref{eq:3} tends to infinity (see
Fig.~\ref{fig:2b}). Also apparent and event horizons arise. The
apparent horizon corresponds to events where world lines
$r=\const$ are horizontal or vertical (see \cite{Zeldovich71} page
85). All space-time is divided into $R$- and $T$-regions (see
\cite{Frolov98,Nov64-1,Nov64-2,Nov01}). In the $R$-regions (lines
"a"\, and "b"), the lines ${r=\const}$ are time-like and signals
can go both to higher and smaller $r$. In the $T$-region (between
the two apparent horizons) the lines ${r=\const}$ are space-like
and a signal can propagate to smaller $r$ only. The structure of
the apparent horizon depends on the value of the flux of the
energy through it. We will analyze it later (see
section~\ref{sec:V}).

\begin{figure*}
\subfigure[Contour lines of the resulting $\Psi$-flux $=
T^\Psi_{vv}-T^\Psi_{uu}$ with a factor of $10^{\frac{1}{2}}$
between lines. The lines "e"\, and "f"\, denotes
$T^\Psi_{vv}-T^\Psi_{uu}= + 10^{-7}$ and $T^\Psi_{vv}-T^\Psi_{uu}=
- 10^{-7}$ respectively, with the resulting absolute flux
increasing towards the $r=0$ singularity. The regions marked "A"\,
and "C"\, marks $ T^\Psi_{vv}-T^\Psi_{uu} > 0$,  "B"\, and "D"\,
marks regions with $ T^\Psi_{vv}-T^\Psi_{uu} < 0$. The thick
dotted line marks the contour line $ T^\Psi_{vv}-T^\Psi_{uu} = 0$
and the fully drawn thick line marks the $r=0$ singularity as in
previous plots as well as the region "X"\, which marks the area
outside of the computational domain.
\label{fig:3c}]{\includegraphics[width=0.49\textwidth]{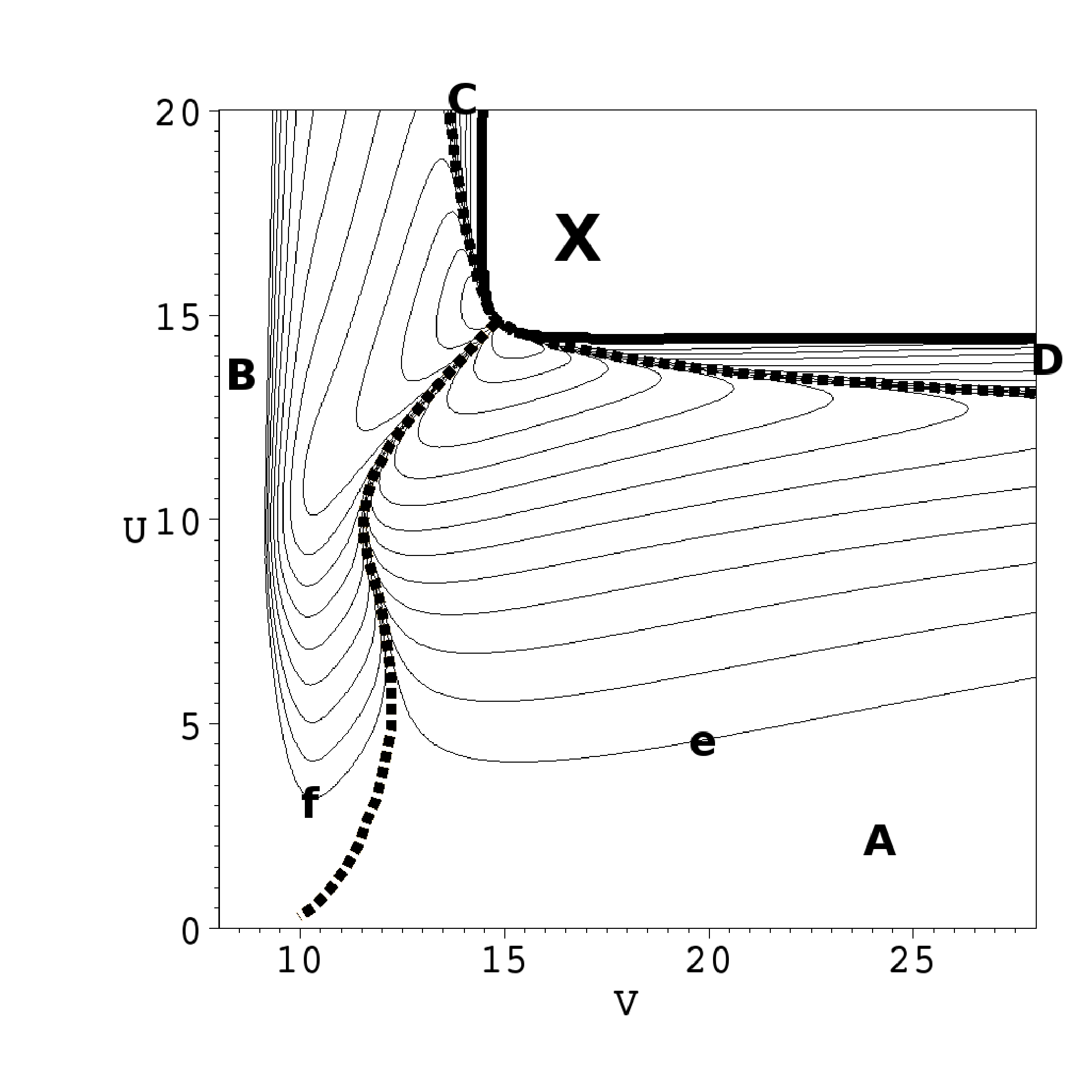}}
\subfigure[Comparison of the $\Phi$ and $\Psi$ fluxes along the
line $u=3$. Line "A"\, marks $|T^\Psi_{vv}-T^\Psi_{uu}|$, line
"B"\, marks $ T^\Phi_{vv} $ and line "C"\, marks $ T^\Phi_{uu}$.
\label{fig:3d}]{\includegraphics[width=0.49\textwidth]{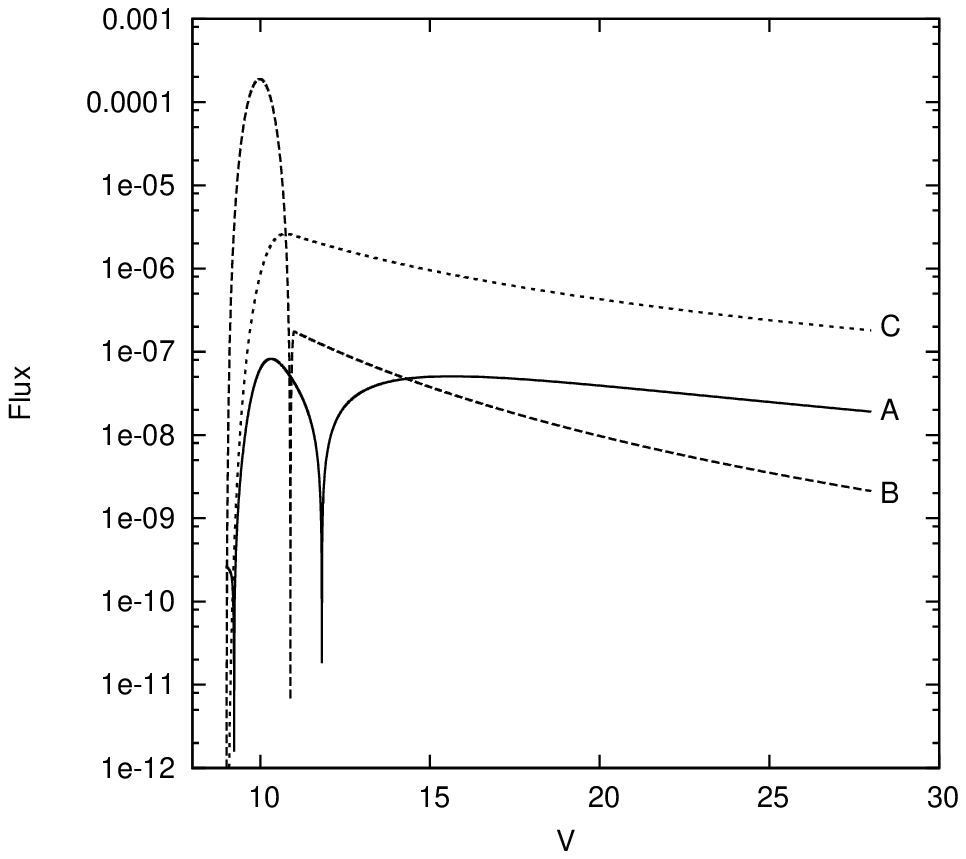}}
\subfigure[Comparison of the $\Phi$ and $\Psi$ fluxes along the
line $u=14$. Line "A"\, marks $|T^\Psi_{vv}-T^\Psi_{uu}|$, line
"B"\, marks $ T^\Phi_{vv} $ and line "C"\, marks $ T^\Phi_{uu}$.
\label{fig:3e}]{\includegraphics[width=0.49\textwidth]{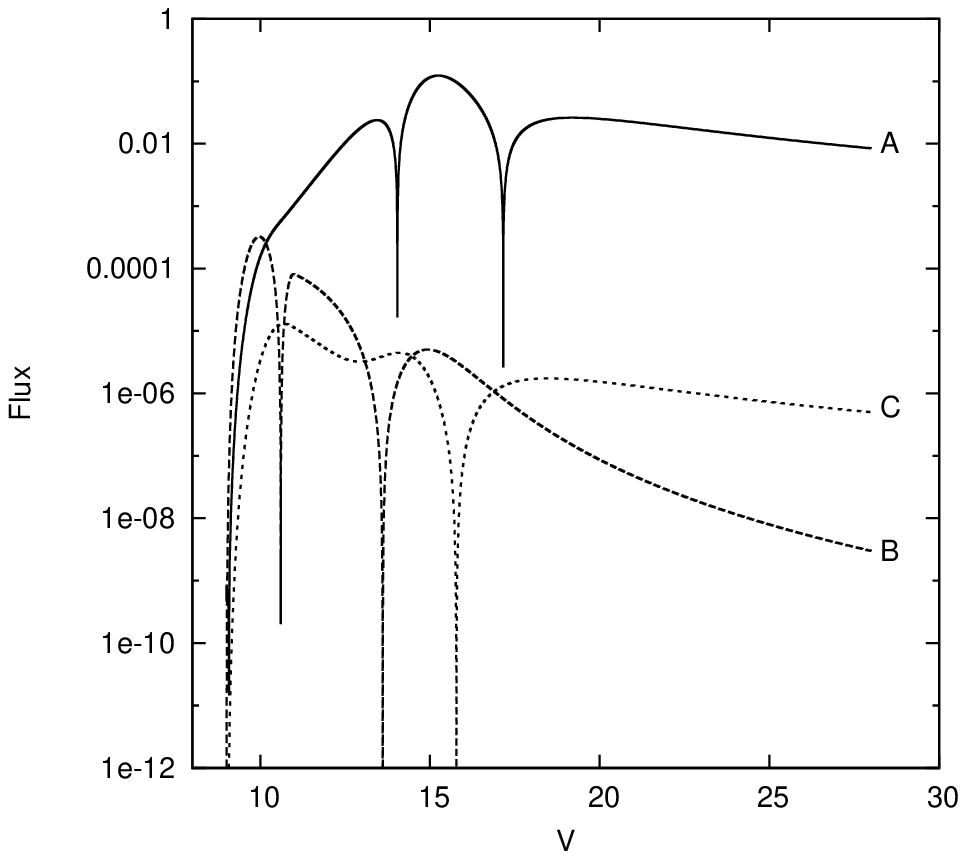}}
\subfigure[Comparison of the $\Phi$ and $\Psi$ fluxes along the
line $u=20$. Line "A"\, marks $|T^\Psi_{vv}-T^\Psi_{uu}|$, line
"B"\, marks $ T^\Phi_{vv} $ and line "C"\, marks $ T^\Phi_{uu}$.
\label{fig:3f}]{\includegraphics[width=0.49\textwidth]{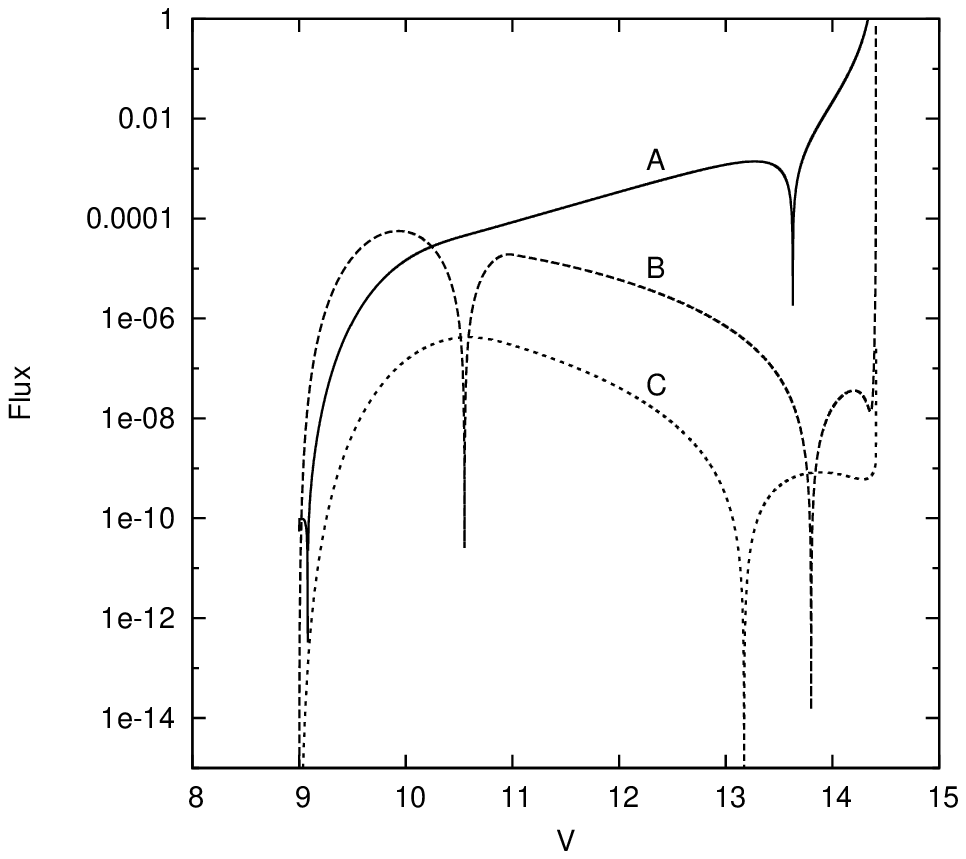}}
\caption{{\label{fig:3_2} Results of the simulation with $v_1 = 9,
v_2 = 11$ and amplitude $A^{\Phi} = 0.01$.}}
\end{figure*}

Now we consider the evolution of the ${\Phi}$- and $\Psi$-fields.
Fig.~\ref{fig:3a} represents the evolution of the flux
${T^\Phi_{vv}}$ of the scalar $\Phi$-energy into the wormhole.
Fig.~\ref{fig:3b} shows the evolution of the ${T^\Phi_{uu}}$ flux
which arises as a result of ${T^\Phi_{vv}}$ being scattered by the
space-time curvature. Fig.~\ref{fig:3c} represents the evolution
of the difference of the ${T^\Psi_{vv}-T^\Psi_{uu}}$ of the fluxes
of the $\Psi$-field in and out of the wormhole. The reason why for
the $\Psi$-field we show the difference of the in and out fluxes
rather than the ${T^\Psi_{vv}}$ and ${T^\Psi_{uu}}$-fluxes
separately, is the fact that even for the static MT-solution
\eqref{eq:lineelement}, \eqref{eq:2}, the values of
${T^\Psi_{vv}}$ and ${T^\Psi_{uu}}$ are not equal to zero (but
equal to each other). So only the difference between them is
important for the dynamics. Finally Fig.~\ref{fig:3d} -
\ref{fig:3f} shows the comparison of all three fluxes at different
slices:
$$u=3,\, 14,\, 20.$$

In Fig.~\ref{fig:3a} and \ref{fig:3b} the initial pulse (between
${9.0<v<11.0}$) and subsequent tails with resonances are clearly
seen. In different regions, $T^\Phi_{uu}$ and $T^\Phi_{vv}$ are
converted into one another due to curvature scattering and
resonances. But at later $u$ (see Fig.~\ref{fig:3e}, \ref{fig:3f})
and for $v>11$ the flux $T^\Psi_{vv}-T^\Psi_{uu}$ of the
$\Psi$-field dominates absolutely (in modulus). Thus for these
regions this flux of the $\Psi$-field determines the dynamics of
the wormhole.

From the Fig.~\ref{fig:2}-\ref{fig:3_2} the general picture of the
evolution of the wormhole under the action of the passage of the
compact $\Phi$-field signal looks like the following. At the
beginning, a rather weak signal produces small perturbations of
the static wormhole. These perturbations trigger the evolution of
the $\Psi$-field that maintains the wormhole. The fluxes of the
$\Psi$-field (in- and out-fluxes) determine the subsequent
evolution of the wormhole and lead to its collapse. At the
beginning of this process the fluxes of the $\Psi$-field are
directed to the throat from both sides of the wormhole (regions A
and B in fig. \ref{fig:3c}). This flux to the throat of the
$\Psi$-field with the negative energy density may lead to the
formation of the negative local effective mass. Remember that
before the beginning of the process the local effective mass was
equal to zero everywhere (no any gravitational forces). Just after
the passage of the perturbing compact ingoing signal of the
$\Phi$-field, the effective mass become positive by the mass of
the ingoing $\Phi$-shell. In the process of the collapse the
fluxes directed in the opposite directions (out of the throat)
arise (regions C and D in fig. \ref{fig:3c}). These fluxes partly
propagate out of the openings of the wormhole and partly propagate
inside of the forming BH. The fluxes of the $\Psi$-field going out
of the wormhole carry away negative energy providing origin of the
positive mass of the forming BH. That is emphasized in papers
\cite{dop-6,dop-7}. Everything looks like a collapse and an
explosion of the $\Psi$-field.

A very important fact is that the compact signal of the
$\Phi$-field propagates through the wormhole from one
asymptotically flat region into another asymptotically flat region
before the beginning of the collapse, this is especially clear
from fig. \ref{fig:3a}. Only a small part of the $\Phi$-signal is
scattered back and into the arising BH. After the BH formation the
propagation of a signal from one asymptotically flat space into
another one is impossible in any direction.

\section{Stronger $\Phi$-signal, horizons and rate of the collapse}
\label{sec:V}

\begin{figure*}
\subfigure[Lines of constant $r$, from $r=0$ to $r=5$ with $\Delta
r = 0.2$ (lines "a"\, marks $r=1.2$ and line "b"\, marks $r=5$).
Thick dotted line marks the position of the apparent horizons.
\label{fig:4a}]{\includegraphics[width=0.49\textwidth]{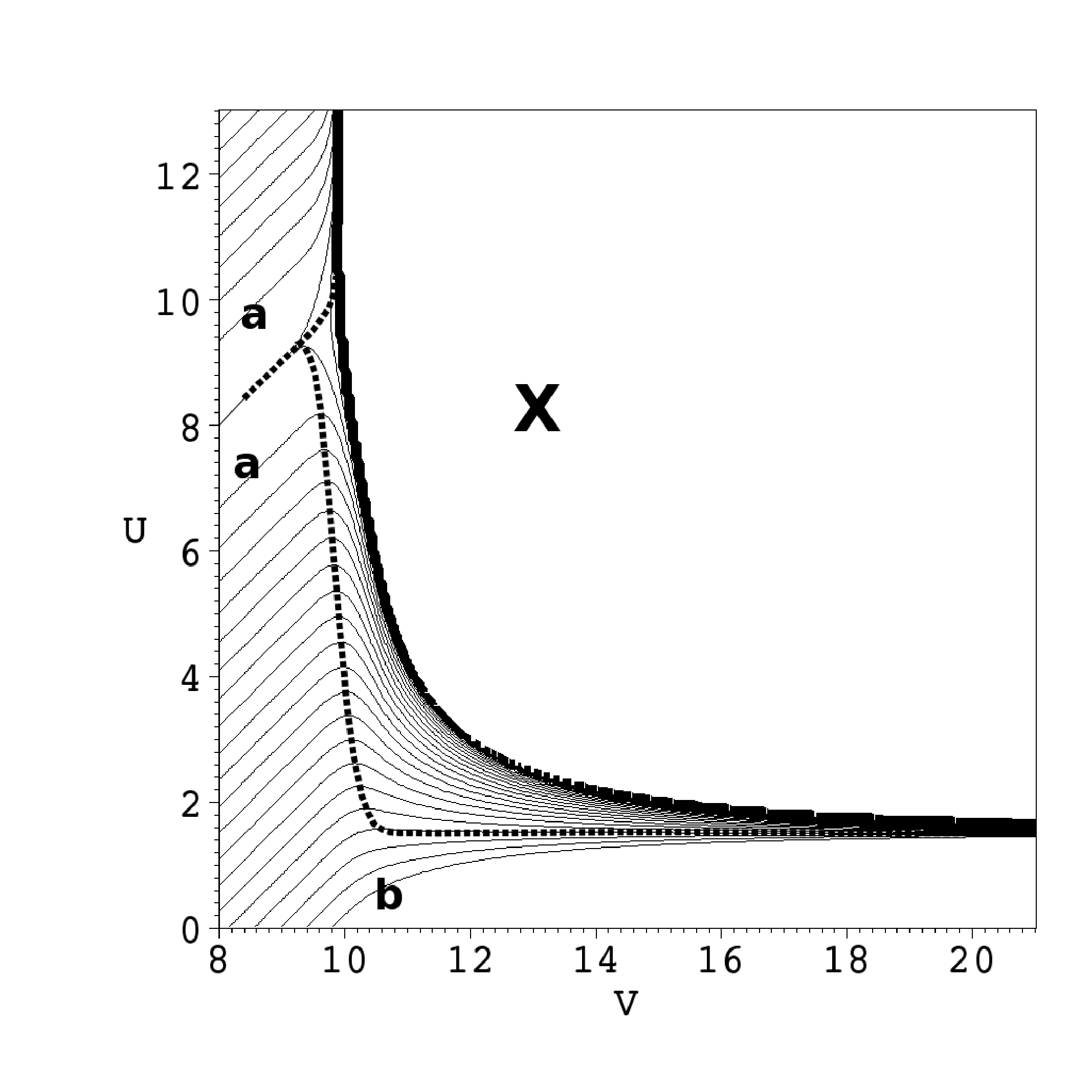}}
\subfigure[Lines of constant ${T^\Phi_{vv}}$. Lines are
logarithmically spaced with a factor $10^{\frac{1}{2}}$ between
lines, line "a"\, marks ${T^\Phi_{vv}}=10^{-0.5}$ and line "b"\,
marks ${T^\Phi_{vv}}=10^{-5}$.
\label{fig:4b}]{\includegraphics[width=0.49\textwidth]{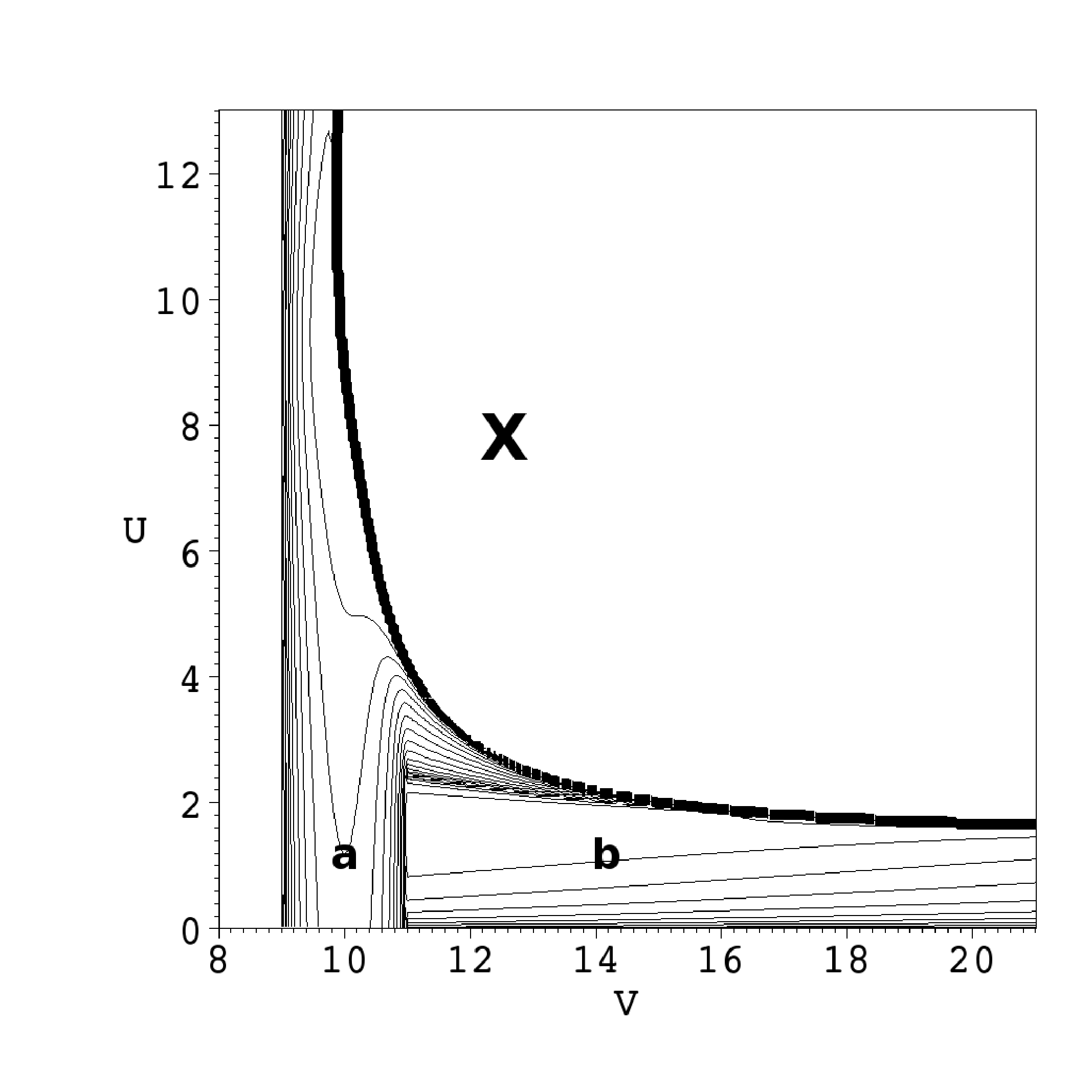}}
\subfigure[Lines of constant ${T^\Phi_{uu}}$. Lines are
logarithmically spaced with a factor $10^{\frac{1}{2}}$ between
lines, line "a"\, marks ${T^\Phi_{vv}}=10^{-2.5}$.
\label{fig:4c}]{\includegraphics[width=0.49\textwidth]{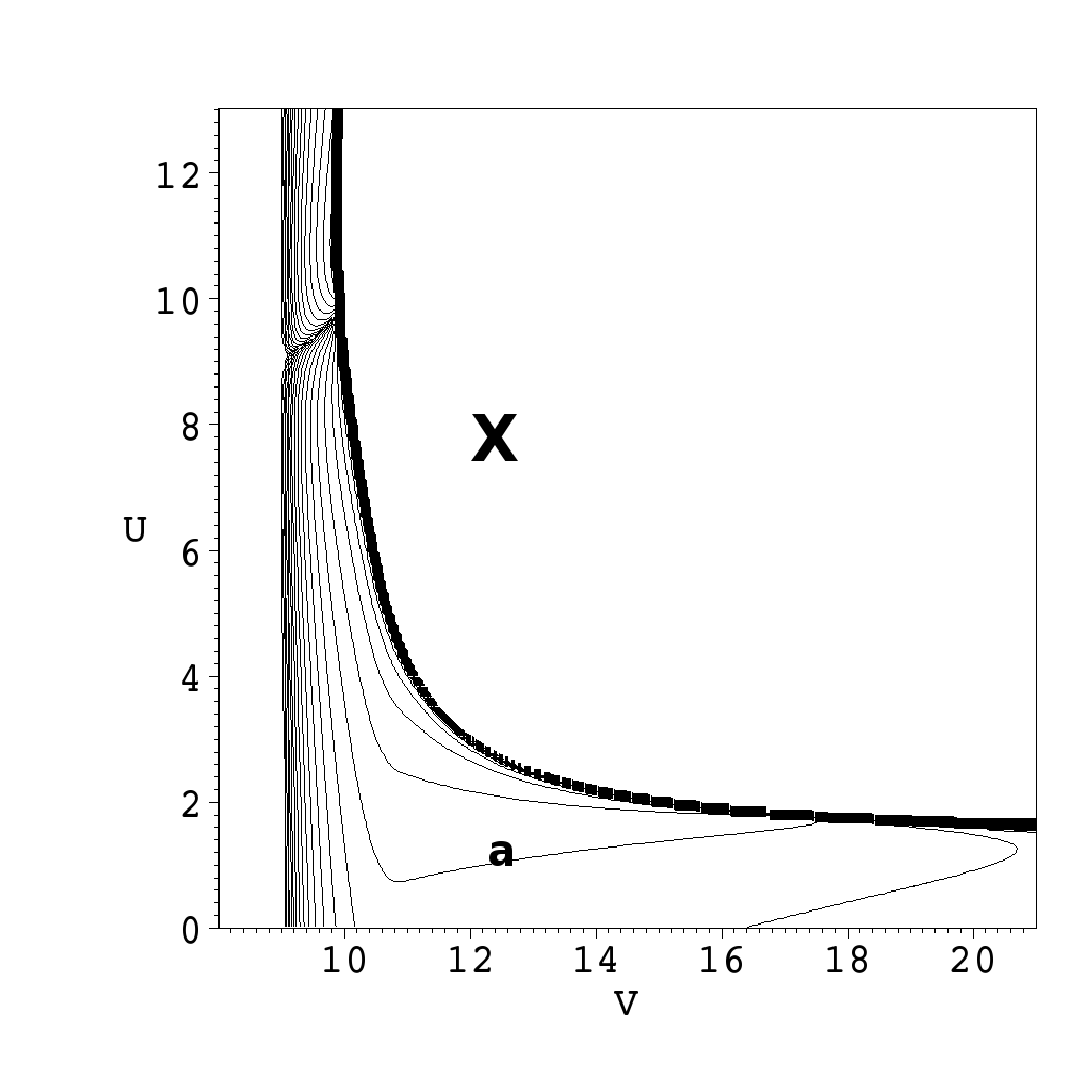}}
\subfigure[Lines of constant $T^\Psi_{vv}-T^\Psi_{uu}$, Region
"A"\, and "C"\, marks regions with $T^\Psi_{vv}-T^\Psi_{uu} > 0$,
"B"\, and "D"\, marks regions with $T^\Psi_{vv}-T^\Psi_{uu} < 0$
and the thick dotted line marks $T^\Psi_{vv}-T^\Psi_{uu} = 0$.
\label{fig:4d}]{\includegraphics[width=0.49\textwidth]{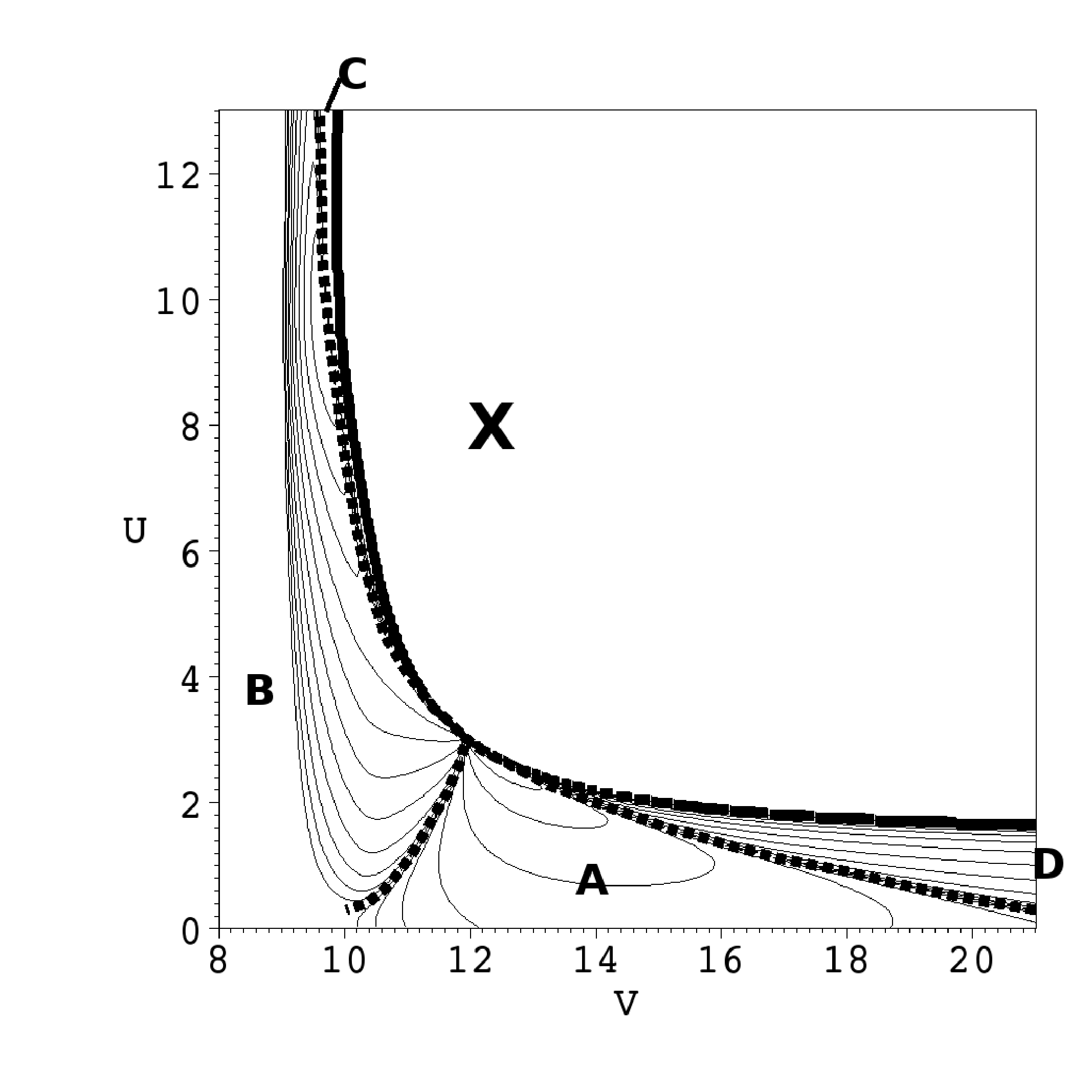}}
\caption{{\label{fig:4} Results for a simulation with $v_1 = 9,
v_2 = 11$ and amplitude $A^{\Phi} = 0.5$. For all figures, the
thick fully drawn line marks the position of the singularity
$r=0$, and the region marked "X"\, marks area outside of the
computational domain. }}
\end{figure*}

Now let us consider the case of an essentially stronger
infalling $\Phi$-signal with $A^{\Phi}=0.5$ (see
Fig.~\ref{fig:4a}-\ref{fig:4d}).

In this case, the collapse of the wormhole arises much faster. The
essential part of the initial $\Phi$-signal comes into the arising
BH, this can be seen from figure \ref{fig:4b}. The picture of the
process is very asymmetrical.

The asymmetry is clearly seen at the picture of the apparent
horizon (figure \ref{fig:4a}). In general, with any strength of the compact
$\Phi$-signal, the apparent horizon arises at the throat just at
the moment of the passage of the $\Phi$-signal (see
Fig.~\ref{fig:2a}). At the beginning the dynamics of the wormhole is
rather slow (if the $\Phi$-signal is weak enough) and two branches
of the apparent horizon are very close from each other.

When the fast collapse begins these branches are going in
different directions.

In the case of a strong initial pulse of the $\Phi$-field
(Fig.~\ref{fig:4a}) the shape of the branch of the apparent horizon
from the side of the coming $\Phi$-field signal is quite different
from the one at the opposite side.

The analogous change of the apparent horizon under the action of
the incoming signal we observe in the charged BH, see
\cite{Hansen05}.

The picture of the evolution of the $\Psi$-field in the case of
the strong signal (see Fig.~\ref{fig:4d}) is qualitatively similar
to the case of weaker signals (Fig.~\ref{fig:3c}).

\begin{figure*}
\subfigure[$r$ as a function of the distance parameter
$L=\sqrt{(v-v_0)^2 + (u-u_0)^2}$ with $v_0=u_0=9$. The curves "a",
"b", "c", "d", "e", "f"\, and "g"\, marks simulations with $A^\Phi
= 0.5, 0.2, 0.1, 0.01, 0.001, 0.00035$ and $0.0001$ respectively.
\label{fig:5a}]{\includegraphics[width=0.49\textwidth]{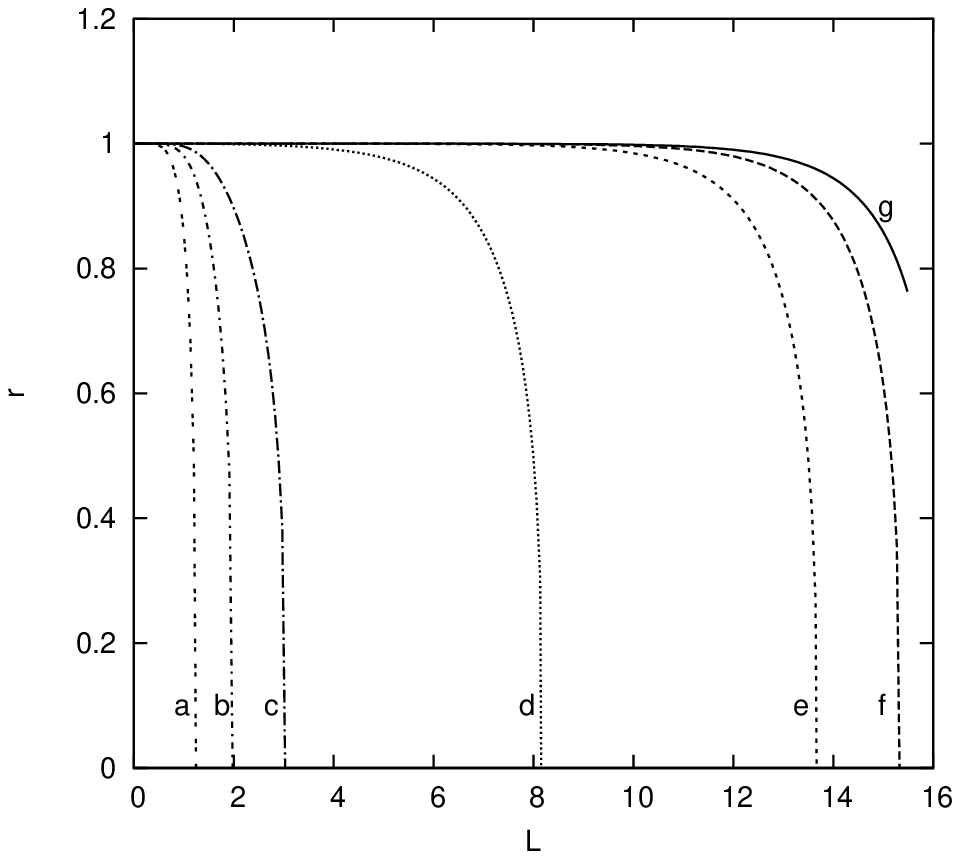}}
\subfigure[Distance parameter $L=\sqrt{(v-v_0)^2 + (u-u_0)^2}$
with $v_0=u_0=9$ measured along $u=v$ from the start of the pulse
to the position of the $r=0$ singularity as a function of scalar
field amplitude $A^\Phi$.
\label{fig:5b}]{\includegraphics[width=0.49\textwidth]{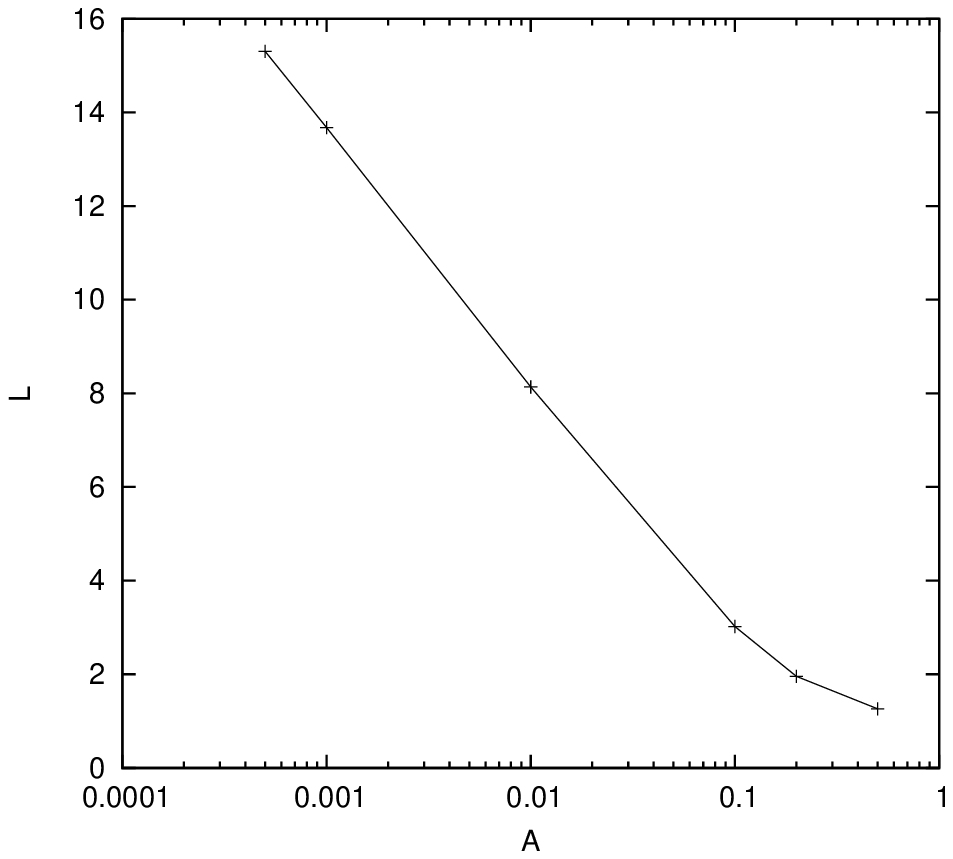}}
\caption{{\label{fig:5} Development of the $r=0$ singularity as a
function of pulse amplitude $A^\Phi$ (with identical pulse widths
of $v_1 = 9, v_2 = 11$).}}
\end{figure*}

About the event horizon. From Fig.~\ref{fig:2a} it is clear that
events close enough to the singularity $r=0$ are
inside of the BH and thus inside of the event horizon. Indeed, the light
signal from these events (vertical and horizontal lines) will come
to the singularity and cannot go to infinity. At Fig.~\ref{fig:2a}
event horizon practically coincide with vertical and horizontal
asymptotes to the singularity $r=0$. The analogous region inside
the BH can be found at the Fig.~\ref{fig:4a}.

Now let us consider the process of the collapse of the throat. In
Fig.~\ref{fig:5a} one can see this process for different amplitudes
$A^{\Phi}$ of the ingoing signal of the $\Phi$-field. It is seen that for
small $A^{\Phi}$ the wormhole practically does not responds to the
$\Phi$-signal during long periods and after that collapses very
fast. It looks like a very nonlinear process.

Fig.~\ref{fig:5b} represents the dependence of the moment of
collapse of the throat on the amplitude $A^{\Phi}$ of the ingoing
$\Phi$-signal. It is seen that for small $A^{\Phi}$ the period between
the passage of the signal and the moment of the collapse can be
very long.

\section{Passage of the $\Psi$-field}
\label{sec:VI}

In this section we briefly consider the passage of the
$\Psi$-field compact pulse through a MT-wormhole. As we told at
the end of section~\ref{sec:I}, the MT wormhole is unstable against
small spherical perturbations. Here we consider not small
perturbations in the form of a compact signal coming from outside.
As we will see, the evolution is in agreement with our conclusion
about instability of the MT wormhole. Now we put the
$\Phi$-field$\equiv 0$ everywhere and model the ingoing
$\Psi$-flux as a perturbation to the background $\Psi^{(MT)}$
field supporting the wormhole:
\begin{equation}
  \label{eq:psi_perturbation}
\Psi (u_0,v) = \Psi^{(MT)}(u_0,v) + \bar\psi (u_0,v)
\end{equation}

The form for the perturbation of the $\Psi$-field into the wormhole is
specified almost the same way as in the case of the $\Phi$-field
in the section \ref{sec:IV}. But now instead of the formula
(\ref{eq:12}) we should write down
\begin{equation}
  \label{eq:14}
\bar\psi_{,v} (u_0 ,v) = A^{\Psi} \sin^2 \left( \pi \frac{v-v_1}{v_2 -
v_1}\right) ,
\end{equation}
All other remarks about the initial conditions from section
\ref{sec:IV} are correct here also. But now there is one new
factor in this case. Namely, this pulse of the $\Psi$-field is an
addition to the background of the $\Psi$-field of the MT solution.
This addition can be taken with the sign plus or minus. This means
that $A^{\Psi}$ can be positive or negative. The evolution depends
critically on this sign.

\begin{figure*}
\subfigure[Lines of constant $r$, from $r=1$ to $r=5$ with $\Delta
r = 0.5$ between lines (line "a"\, marks $r=1.5$ and line "b"\,
marks $r=11$,
\label{fig:6a}]{\includegraphics[width=0.49\textwidth]{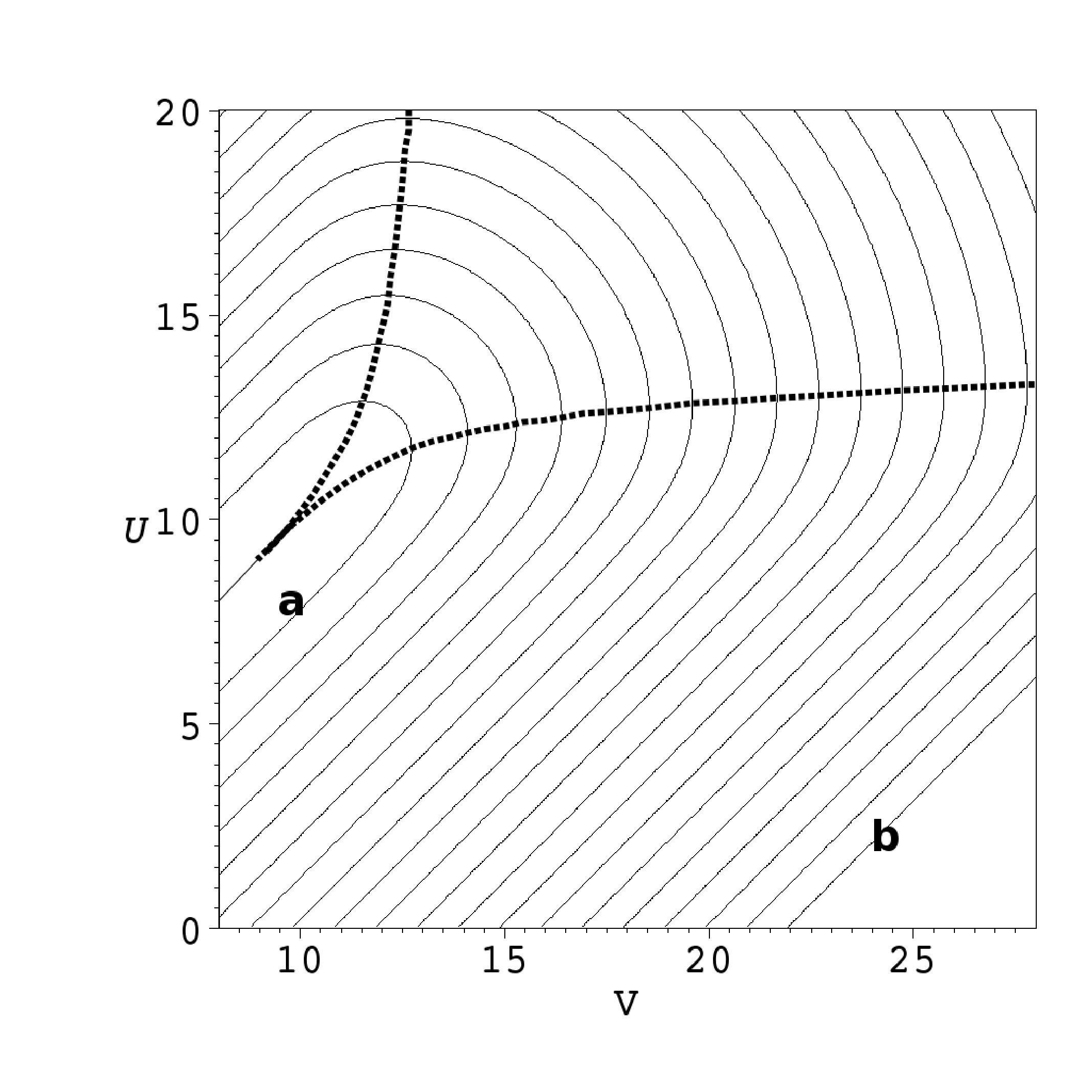}}
\subfigure[Lines of constant $T^\Psi_{vv}-T^\Psi_{uu}$, Region
"a"\, marks region with $T^\Psi_{vv}-T^\Psi_{uu} < 0$ "b"\, marks
region with $T^\Psi_{vv}-T^\Psi_{uu} > 0$ and the thick dotted
line marks $T^\Psi_{vv}-T^\Psi_{uu} = 0$.
\label{fig:6b}]{\includegraphics[width=0.49\textwidth]{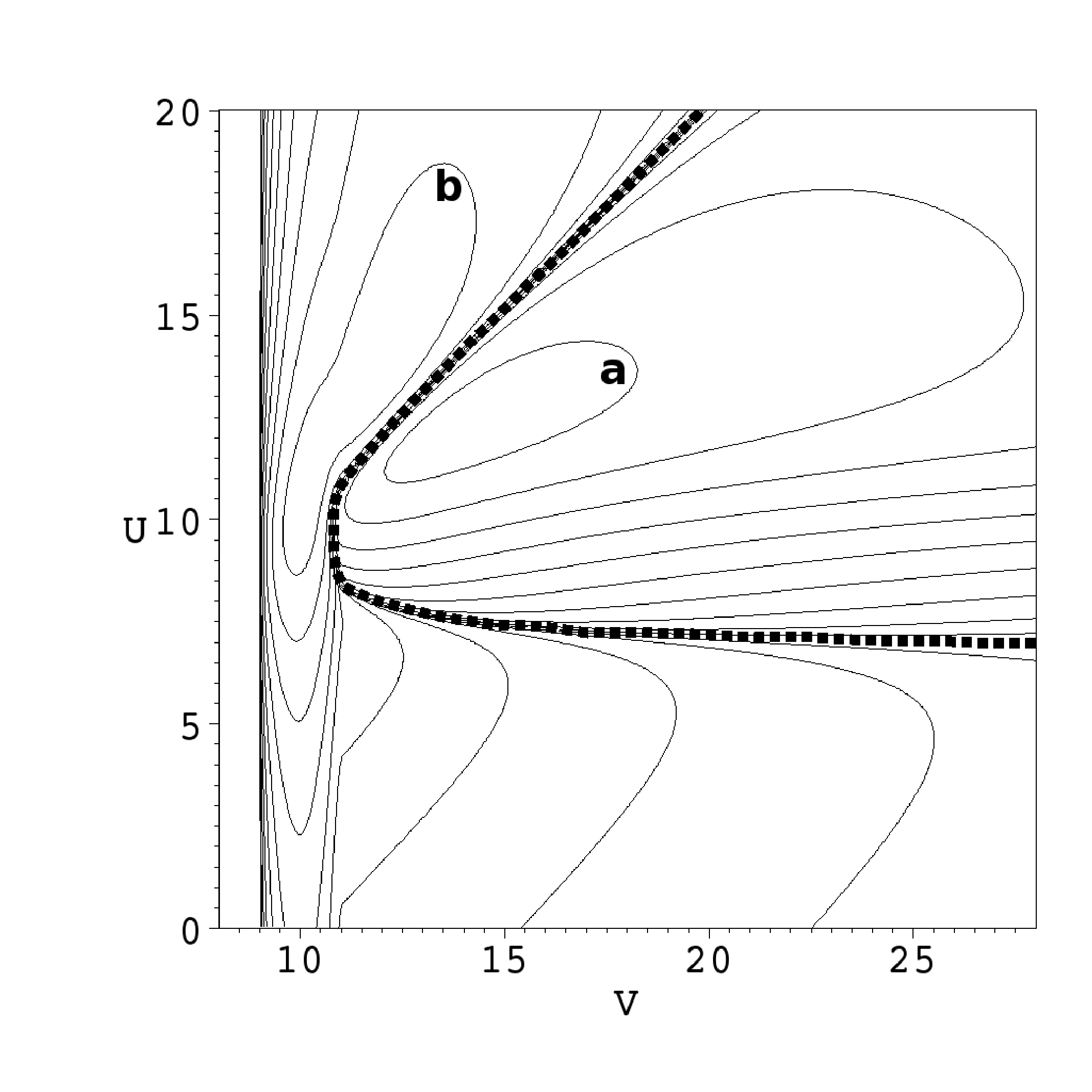}}
\caption{{\label{fig:6} Results for a simulation with $v_1 = 9,
v_2 = 11$ and $A^{\Psi} = +0.01$ (${\Phi} = 0$ everywhere).}}
\end{figure*}

Fig.~\ref{fig:6} represent the case of the expression
(\ref{eq:14}) with the sign $A^{\Psi} > 0$. The wormhole expands
(Fig.~\ref{fig:6a}) and the size of the throat becomes larger and
larger. The physical reason for this expansion is clear, the additional
amount of the $\Psi$-field with the negative energy density
creates gravitational repulsion. This gives the initial push. The
expansion at the late $v$-parameter looks like an almost
"inertial".

Between two branches of the apparent horizon any test particle or
radiation can move to greater and greater $r$ only. This is a so
called expanding $T_{+}$ region, see
\cite{Frolov98,witten2001,chicago,texbook,Huetal2000}.

Fig.~\ref{fig:6b} represent the fluxes of the $\Psi$-field out of
the throat to greater $r$.

\begin{figure*}
\subfigure[Lines of constant $r$, from $r=1$ to $r=10$ with
spacing $\Delta r = 0.5$ between lines, lines "a"\, marks $r=1.5$,
line "b"\, marks $r=10$. Thick dotted lines marks positions of
apparent horizons.
\label{fig:7a}]{\includegraphics[width=0.49\textwidth]{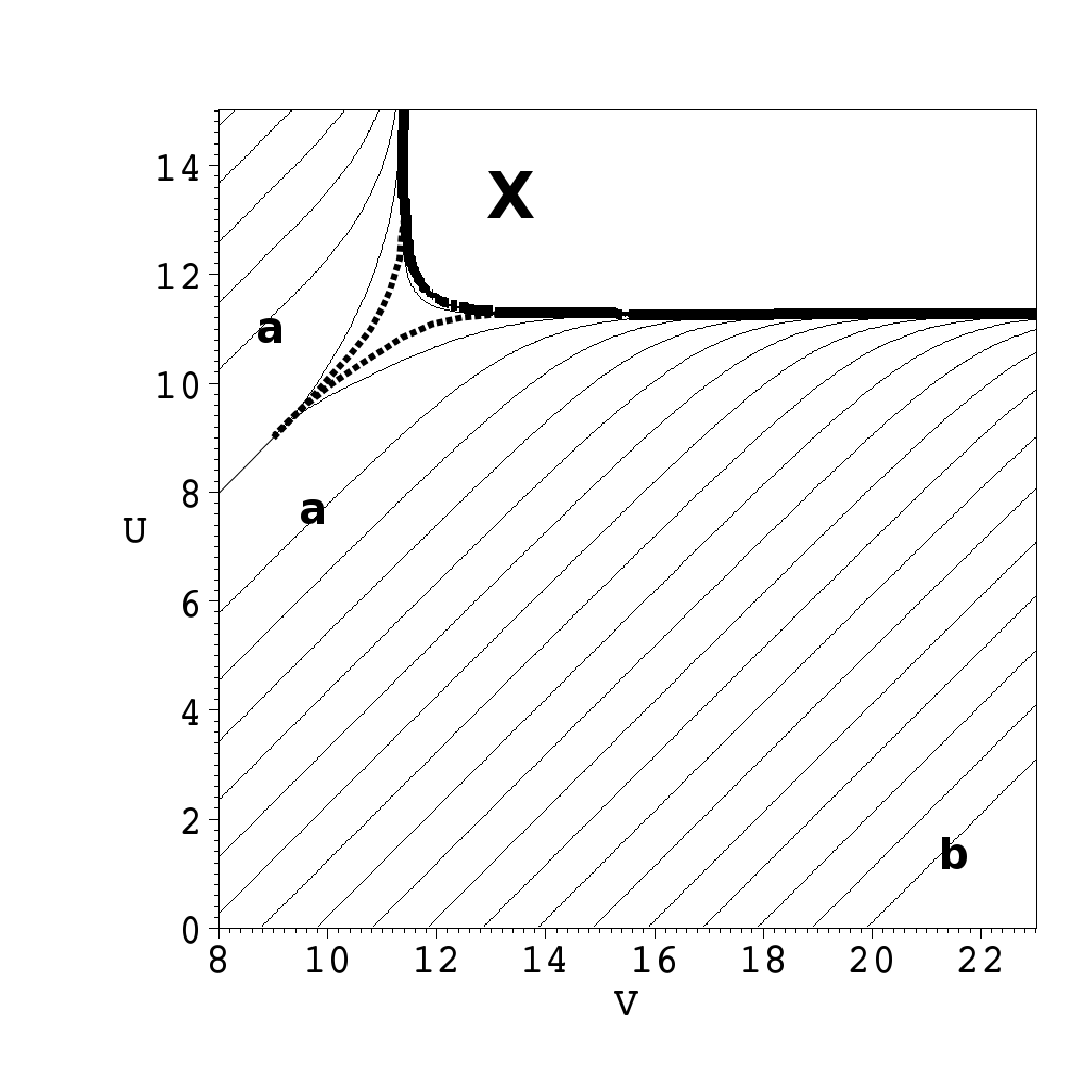}}
\subfigure[Lines of constant $T^\Psi_{vv}-T^\Psi_{uu}$, Regions
"A"\, and "C"\, marks regions with $T^\Psi_{vv}-T^\Psi_{uu} > 0$,
"B"\, and "D"\, marks regions with $T^\Psi_{vv}-T^\Psi_{uu} < 0$
and the thick dotted line marks $T^\Psi_{vv}-T^\Psi_{uu} =
0$.\label{fig:7b}]{\includegraphics[width=0.49\textwidth]{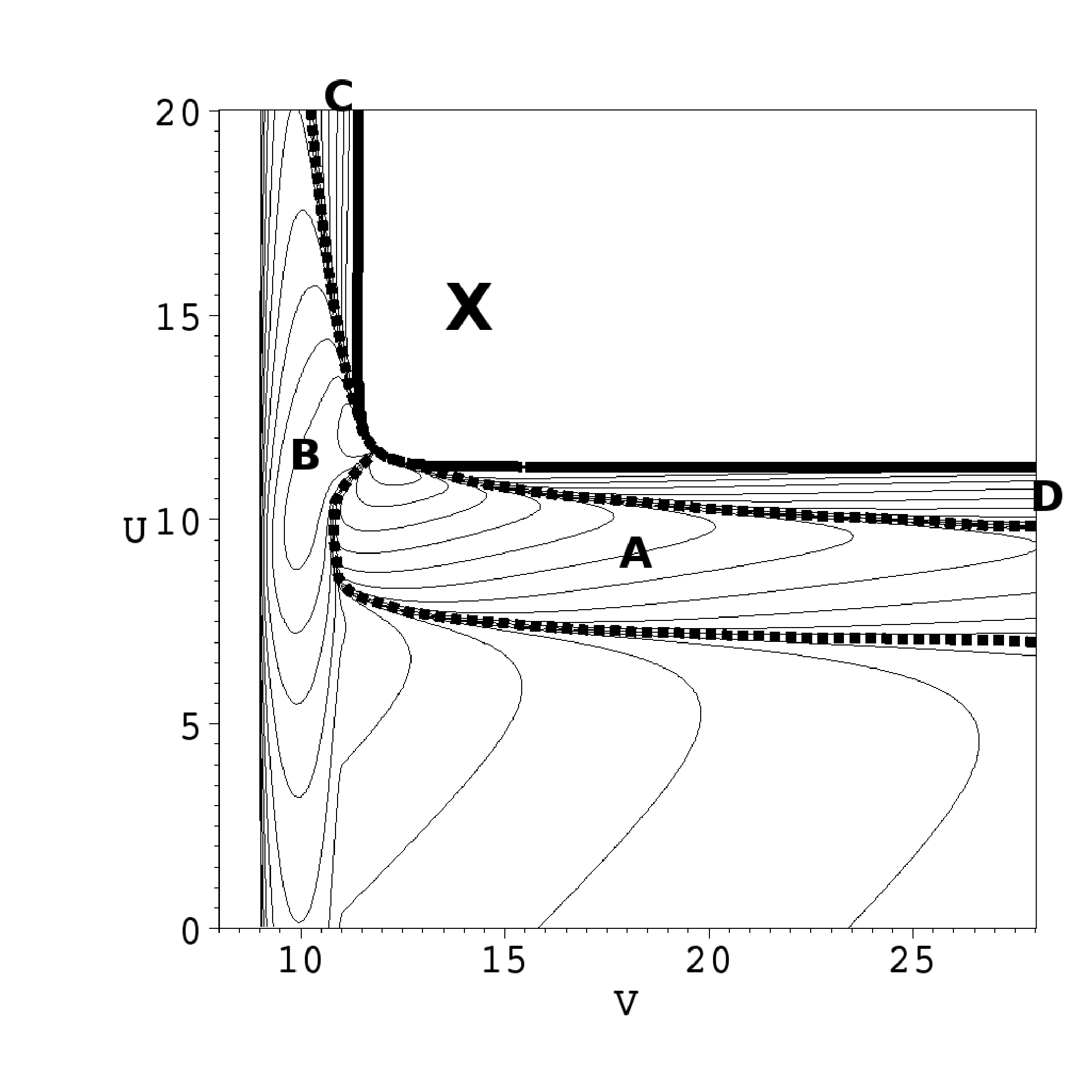}}
\caption{{\label{fig:7} Results for a simulation with $v_1 = 9,
v_2 = 11$ and $A^{\bar\Psi} = -0.01$ (${\Phi} = 0$ everywhere). As
previously, the thick fully drawn line marks the position of $r=0$
and the region marked "X"\, indicates the region beyond our
computational domain.}}
\end{figure*}

The case of the expression (\ref{eq:14}) with the sign $A^{\Psi} <
0$ is represented in Fig.~\ref{fig:7}. Now there is a deficit of
the $\Psi$-field as compared with the static solution
(\ref{eq:lineelement}), (\ref{eq:2}). It leads to collapse of the
wormhole. Qualitatively the process of the collapse is the same as
it was described in section \ref{sec:IV}.

\section{Conclusions}
\label{Conclusion}

In this paper we investigated numerically the process of the
passage of the radiation pulse through a wormhole and the
subsequent evolution of the wormhole as respond to a gravitational
action of this signal. It is the continuation of the
investigations started in the papers \cite{dop-6,dop-7,dop-8}. In
our work we focus mainly on the evolution of the fields $\Phi$ and
$\Psi$. We use the MT solution
[see~\eqref{eq:lineelement}-\eqref{eq:2}] of the Einstein
equations as a model of an initial static wormhole.

The spherically symmetrical radiation pulse was modeled by the
self-gravitating, minimally coupled, massless scalar fields $\Phi$
and $\Psi$ [see eqs \eqref{eq:12}-\eqref{eq:14}]. We created and
tested a numerical code which is stable and second order accurate.
For our computations we used an adaptive mesh refinement approach
in both ingoing $u$ and outgoing $v$ directions. We have
demonstrated that if an amplitude of the $\Phi$-field, $A^{\Phi}$,
is small enough, then just after the propagation of it through the
throat the perturbations of the static wormhole are small. The
compact signal of the $\Phi$-field propagates through the wormhole
from one asymptotically flat region into another one. The
scattering of the signal is small also. But the small
perturbations trigger the evolution of the $\Psi$-field which
maintains the wormhole. The fluxes of the $\Psi$-field (in- and
out-fluxes) determine the subsequent evolution of the wormhole and
lead to its collapse and a BH arises.

After the BH formation the propagation of a signal from one
asymptotically flat space into another one is impossible in any
direction.

In the case of the strong enough $\Phi$-signal in-falling into the
wormhole, the collapse of the wormhole arises much faster. The
essential part of the initial $\Phi$-signal falls into an arising BH.

We analyzed the dependence of the period between the passage of the
signal through the throat and the moment of the collapse of the
throat in $r=0$ on the amplitude $A^{\Phi}$. For smaller and smaller
amplitude $A^{\Phi}$, this period becomes longer and longer.

Note, that the collapse and the BH formation after the passage of
the $\Phi$-signal through the wormhole is typical for this special
type of the wormholes supported by a $\Psi$-field. Other
types of wormholes may behave quite different. We will
investigate such cases in another paper.

Finally we investigated the passage of a compact signal of the
$\Psi$-field through the wormhole and evolution of the $\Psi$
field in this case.

As we emphasized above our work continues the work
\cite{dop-6,dop-7}. We investigate mainly the evolution of the
$\Phi$ and $\Psi$ fields during the nonlinear evolution of the
wormhole. On the other hand we want to emphasize the difference in
our approach from the approach in the paper \cite{dop-7}. In
\cite{dop-7} the authors formulate the initial condition for
perturbations in such way that it corresponds to two fluxes of the
scalar fields which propagate in positive and negative directions.
The center of the perturbations is at the throat of the wormhole.
Our formulation of the initial condition corresponds to a compact
signal coming into the wormhole from the right-hand asymptotically
flat region. This relates our approach with possible applications
to wormhole astrophysics.

\section{Acknowledgements}
\label{Acknowledgements}

The authors thanks Pavel Ivanov for useful discussions and Sean Hayward, Hisaaki Shinkai, Jose
Antonio Gonzalez, Francisco Guzman and Olivier Sarbach, who called our attention to their important papers.

This work was supported in part by the JSPS Postdoctoral Fellowship For Foreign Researchers, the Grant-in-Aid for Scientific Research Fund of the JSPS ${(19-07795)}$, Russian Foundation for Basic Research
(project codes: $07-02-01128-a$, $08-02-00090-a$,
$08-02-00159-a$), scientific schools: $NSh-626.2008.2$,
$Sh-2469.2008.2$ and by the program {\it Origin and Evolution of
Stars and Galaxies 2008} of Russian Academy of Sciences.

The authors thank the Niels Bohr Institute for hospitality during
their visit.

\appendix

\section{Instability}
\label{sec:app0}

For the spherical metric
\begin{eqnarray}
ds^2= -d\tau^2+e^\lambda dR^2+ r^2\, d\Omega^2 ,\,\,\,
r^2=e^\eta(R^2+Q^2). \label{metric}
\end{eqnarray}
the Einstein's tensor $G^i_k$ is:
\begin{eqnarray}
G_\tau^\tau=e^{-\lambda}\left[\eta_{,_{RR}}+\frac{
\eta_{,_R}(3\eta-2\lambda)_{,_R}}{4}\right.+ \label{00}
\end{eqnarray}
\[
+\frac{R(3\eta- \lambda)_{,_R}}{{Y}^2}+\left.
\frac{2Q^2+R^2}{{Y}^4}\right] -\frac{e^{-\eta}}
{{Y}^2}-\frac{2\lambda_{,\tau}\eta_{,\tau}+\eta_{,\tau}^2}{4}
\]
\begin{eqnarray}
G_R^R=e^{-\lambda}\left[\frac{R^2}{{Y}^4}+
\frac{\eta_{,_R}^2}{4}+\frac{R\eta_{,_R}}{{Y}^2}\right]-
\label{11} \end{eqnarray}
\[
-\frac{e^{-\eta}}{{Y}^2}-\eta_{,\tau\tau}-\frac{3}{4}\eta_{,\tau}^2
\]
\begin{eqnarray}
G_\theta^\theta=e^{-\lambda}\left[\frac{Q^2}
{{Y}^4}+\frac{\eta_{,_{RR}}+\eta_{,_R}(\eta-\lambda)_{,_R}}
{4}+\right. \label{22}\end{eqnarray}
\[
\left. +\frac{R(2\eta-\lambda)_{,_R}}{2{Y}^2}\right]
-\frac{\lambda_{,\tau\tau}+\eta_{,\tau\tau}}{2}-\frac{\lambda_{,\tau}^2
+\eta_{,\tau}^2+\eta_{,\tau}\lambda_{,\tau}}{4}
\]
\begin{eqnarray}
G_\tau^R=-e^{-\lambda}\left[\eta_{,\tau_R}+\frac{(\eta
-\lambda)_{,\tau}}{2}\left(\eta_{,_R}+\frac{2R}{{Y}^2}\right)\right]
\label{01}
\end{eqnarray}
Here ${Y}^2=R^2+Q^2$.

The equation for the scalar field is
\begin{eqnarray}
\Psi^{;i}_{;i}=\frac{1}{\sqrt{-g}}\frac{\partial}{\partial
x^i}\left(\sqrt{-g}g^{ik}\frac{\partial\Psi}{\partial x^k}
\right)=0 \label{psi}
\end{eqnarray}
Or:
\begin{eqnarray}
\left[\exp\left(\eta+\frac{\lambda}{2}\right)\Psi_{,\tau}
\right]_{,\tau}=\frac{1}{{Y}^2}\left[\exp\left(\eta-\frac{\lambda}
{2}\right){Y}^2\Psi_{,_R}\right]_{,_R} \label{psi-2}
\end{eqnarray}

The energy-momentum tensor $T^i_k$ is:
\begin{eqnarray}
T_\tau^\tau=\frac{1}{8\pi}\left[\Psi^2_{,\tau}+e^{-\lambda}\Psi^2_{,_R}
\right] = -T_R^R \label{tik}\, ,
\end{eqnarray}
\[
T_\theta^\theta=-\frac{1}{8\pi}\left[\Psi^2_{,\tau}-e^{-\lambda}\Psi^2_{,_R}
\right],\quad
T_\tau^R=-\frac{1}{4\pi}e^{-\lambda}\Psi_{,\tau}\Psi_{,_R}
\]
The static MT solution is
\begin{eqnarray}
\eta=0,\quad \lambda
=0,\quad \Psi_{,_R}=\frac{\pm Q}{{Y}^2} \label{Psi}
\end{eqnarray}
\begin{eqnarray}
T_\tau^\tau=-T_R^R=T_\theta^\theta=Q^2/\left(8\pi {Y}^4\right)
\label{MT}
\end{eqnarray}

\subsection{Perturbations}
\label{sec:app0-1}

Further on we will use dimensionless coordinates $x\equiv R/Q$,
${t}\equiv c\tau/Q$ and will consider linearized Einstein's
equations: ${G^i_k=8\pi T^i_k}$. Perturbed components of the energy-momentum
tensor are dimensionless. From (\ref{00}-\ref{tik}) for
${\eta\ll 1}$, ${\lambda\ll 1}$ we get for perturbations:
\begin{eqnarray}
\delta T_{t}^{t}=-\delta T_R^R=\delta T_\theta^\theta=
\frac{\Psi_{,x}}{8\pi}[2\bar\psi_{,x}-\lambda\Psi_{,x}]
\label{prt}
\end{eqnarray}
\begin{eqnarray}
\delta T_{t}^R=-\frac{1}{4\pi}\Psi_{,_R}\bar\psi_{,{t}}
\label{d01}
\end{eqnarray}
\begin{eqnarray}
Q^2\delta G_{t}^R=-\eta_{,{t} x}-\frac{x(\eta-
\lambda)_{,{t}}}{x^2+1} \label{d01-2}
\end{eqnarray}
\begin{eqnarray}
Q^2\delta G_{t}^{t}= \label{d00}
\end{eqnarray}
\[
=\eta_{,xx}+\frac{3x\eta_{,x}+\eta}
{1+x^2}-\frac{x\lambda_{,x}}{x^2+1}-\frac{x^2+2}
{(x^2+1)^2}\lambda
\]
\begin{eqnarray}
Q^2\delta G_R^R=\frac{x\eta_{,x}+\eta}{x^2+1}
-\eta_{,{t}{t}}-\frac{x^2\lambda}{(x^2+1)^2} \label{d11}
\end{eqnarray}
\begin{eqnarray}
Q^2\delta G_\theta^\theta=\label{d22}
\end{eqnarray}
\[
=\frac{\eta_{,xx}-\lambda_{,{t}{t}}
-\eta_{,{t}{t}}}{2}-\frac{x(\lambda_{,x}-2\eta_{,x})}
{2(x^2+1)}-\frac{\lambda}{(x^2+1)^2}
\]

From the condition $\delta G_{t}^{t}=-\delta G_R^R$ we get
\begin{eqnarray}
\eta_{,{t}{t}}=\eta_{,xx}+\frac{x(4\eta-\lambda)_{,x}}{1+x^2}+
2\frac{\eta-\lambda}{1+x^2} \label{f-nu}
\end{eqnarray}
and the condition $\delta G_{t}^{t}=\delta G_\theta^\theta$ leads
to equation
\begin{eqnarray}
\lambda_{,{t}{t}}+\eta_{,{t}{t}}= \label{eq1}
\end{eqnarray}
\[
=-\left[\eta_{,xx}+\frac{x(4\eta
-\lambda)_{,x}}{1+x^2}+2\frac{\eta-\lambda}{1+x^2}\right]=-\eta_{,{t}{t}}
\]
and using one of the solution (\ref{eq1}) we get
\begin{eqnarray}
\lambda=-2\eta,\quad \eta_{,{t}{t}}=\eta_{,xx}+\frac{6x}
{1+x^2}\eta_{,x}+\frac{6\eta}{1+x^2} \label{eq2}
\end{eqnarray}

For $\eta\propto exp(i\omega t)$ we get
\begin{eqnarray}
\eta_{,xx}+\frac{6\,x}{x^2+1}\eta_{,x}+\left[\omega^2+
\frac{6}{x^2+1}\right]\eta=0 \label{omega1}
\end{eqnarray}
\begin{eqnarray}
\eta=-\frac{\lambda}{2}=\frac{f}{(1+x^2)^{3/2}},\quad
\bar\psi=\frac{f_{,x}}{2\sqrt{1+x^2}}\,. \label{sig-psi}
\end{eqnarray}
\begin{eqnarray}
f_{,xx}+\left[\omega^2+\frac{3}{(1+x^2)^2}\right]f=0 \label{ff}
\end{eqnarray}
As is seen from equations (\ref{omega1})-(\ref{ff}) functions
$\eta(x)$, $f(x)$ and $\bar\psi(x)$ include even and odd modes.
This means that for the even mode the function $\bar\psi(x)$ is
extreme at $x=0$ while for the odd mode $\bar\psi(0)=0$.

The Eq. (\ref{ff}) is identical to the one--dimensional
Schr\"odinger's equation with the energy $E=\omega^2$ and the
potential box
\begin{eqnarray}
U=\frac{-3}{(1+x^2)^{2}},\quad I=\int\limits_{-\infty}^\infty
Udx=-1.5\pi\approx -4.7 \label{pot}
\end{eqnarray}
For such equations the general structure of solution is well
known. For example, for $\omega^2>0$ we have,
\begin{eqnarray}
f=e^{i\omega t}\left\{
\begin{array}{cc}
c_1 e^{i\omega x}& \,{\rm for}\, x\ll -1\cr c_2 e^{-i\omega x}+c_3
 e^{i\omega x}& {\rm for} \, x\gg 1\cr
\end{array}
\right. \label{rom}
\end{eqnarray}
what are the incident and reflected waves for ${x\gg 1}$ and the
traveled wave for ${x\ll -1}$.

For the potential (\ref{pot}) there is at least one solution with
$\omega^2=-\alpha^2$, $3\geq\alpha^2\geq 0$ and for finite $f(x)$
at $x\rightarrow \pm\infty$ we expect to get
\begin{eqnarray}
f\sim e^{\alpha t}\left\{
\begin{array}{cc}
c_1\exp(\alpha x)& {\rm for}\, x\ll -1\qquad\cr
c_2\sin(\sqrt{3-\alpha^2}x+\phi)& {\rm for}\, -1\ll x\ll 1\cr
c_3\exp(-\alpha x)& {\rm for} \,x\gg 1\qquad\quad\,\cr
\end{array}
\right. \label{iom}
\end{eqnarray}
with $\phi=const$ and a discrete set of $\alpha=\alpha_i, \,i=1,2,
...$ and $\sqrt{3}\geq\alpha_i\geq 0$.

To estimate the expected principal values $\alpha$ we can use
solutions of the Schr\"odinger's equation with two similar
potential boxes, namely,
\begin{eqnarray}
U=U_1=\left\{
\begin{array}{cc}
\,\,\,\,\,0& {\rm for}\, x\leq -1\cr -3& \qquad\,\,{\rm for}
\,-1\leq x\leq 1\cr \,\,\,\,\,0& \,\,{\rm for}\, x\geq 1\quad\cr
\end{array}
\right. \label{box1}
\end{eqnarray}
and
\begin{eqnarray}
U=U_2=-3{\rm ch}^{-2}x,\quad -\infty\leq x\leq \infty\,.
\label{box2}
\end{eqnarray}
For both potentials (\ref{box1}\,\&\,\ref{box2}) we have
$I=\int_{-\infty}^\infty Udx=-6$ what is close to (\ref{pot}). For
the potential (\ref{box1}) the principal value is
$\alpha=\alpha_1\approx 1.4$ and for the potential (\ref{box2}) it
is $\alpha=\alpha_2=(\sqrt{13}-1)/2 \approx 1.3$. These results
indicate that for Eq. (\ref{ff}) we can expect to get also the
principal value $\alpha\approx 1.3 - 1.4$. In the case the growing
solution of Eq. (\ref{ff}) is located at $|x|\leq x_b\sim 0.6$.

This means that there is an instability in the central region of
the wormhole ($|R|\leq x_bQ$) while at $|R| \geq x_bQ$ the
corresponding function $f$ is exponentially small. These results
demonstrate the instability of the MT solution in respect to the
small perturbations of the field $\Psi$.

\section{Description of the code}
\label{sec:appA}

In this appendix we describe our numerical code used to obtain the
results presented in the paper.

To numerically integrate the evolution equations, eqs.
\eqref{eq:9}-\eqref{eq:11b}, throughout some computational domain,
we have created a new numerical code. The code has adaptive mesh
refinement (AMR) capabilities in both in- and outgoing directions
(contrary to our previous code, which had only AMR along the
ingoing $u$-direction \cite{Hansen05}). The numerical scheme used
to evolve the computational cells is identical to the one we used
in \cite{Hansen05} (which evolves the unknown variables with
second order accuracy), while the AMR algorithm in our code is
largely based upon the design of the algorithm presented in
\cite{Pretorius04}.

A key feature of this AMR algorithm is the usage of a self-shadow
hierarchy to compute the truncation error estimates which
determines whether or not a computational cell should be split to
a higher level. A self-shadow hierarchy works by always
simultaneously evolving two grids, one twice the resolution of the
other. Wherever the computational points in two grids coincide,
it is possible to estimate the local truncation error (TE) by
calculating the difference between the twos solutions. This TE estimate is
then measured against some limit $TE_{max}$ and if the local
truncation error is larger then that allowed, the computational
cell is split to a higher level.

However, the point-wise computation of the TE as described above
is, in general, not the optimal way of computing the TE as the
solutions to the wave-like finite-difference equations is in
general oscillatory in nature and will tend to go to zero at
certain points within the computational domain \cite{Pretorius04}.
Therefore, in practice, when calculating the TE we average the
point-wise TE over several cells to the past of
that point. Where the splitting structure allow it, we use a total
of nine points to the causal past of a point to calculate the TE which determines
whether a cell should be split or not. Furthermore, we calculate
the TE for all four dynamic variables, if any one of these become
greater then our accepted limit $TE_{max}$, the cell is split to a
higher level.

When splitting a cell to a finer level, we need to perform an
interpolation of the cells on the more coarse level in order to
obtain initial data at the refinement boundaries for evolving the
cells on the finer level (see \cite{Hansen05} and
\cite{Pretorius04} for details). If a sufficient number of
adjacent points are available on the coarse parent level we use
cubic Lagrange interpolating polynomials, otherwise we use linear
interpolation.

\section{Testing of the code}
\label{sec:appB}

We have tested our code and found it to be stable and second-order
accurate. As a first, most basic test, we have verified that the
code does indeed converge to the static MT solution.

In this section, we demonstrate that the code is also converging
and second order accurate when including the effects of a
non-trivial infalling $\Phi$ field. The initial conditions for the
tests in this section are similar to those described in section
\ref{sec:IV}, i.e. the initial conditions for the $\Psi$ field is
set to the exact MT solution, and the form of the infalling $\Phi$
field is modeled after eq. \eqref{eq:12} with $v_1 =9, v_2 = 11$
and $A^{\Phi} = 0.0035$. The strength of this $\Phi$ pulse is such
that the wormhole begins its contraction and almost, but not
quite, forms a black hole within our computational domain. Our
computational domain is, as for all simulations in this paper, in
the range $v=[8,28]$ and $u=[0,20]$.

\subsection{Non-AMR convergence tests}
\begin{figure}
\includegraphics[width=0.49\textwidth]{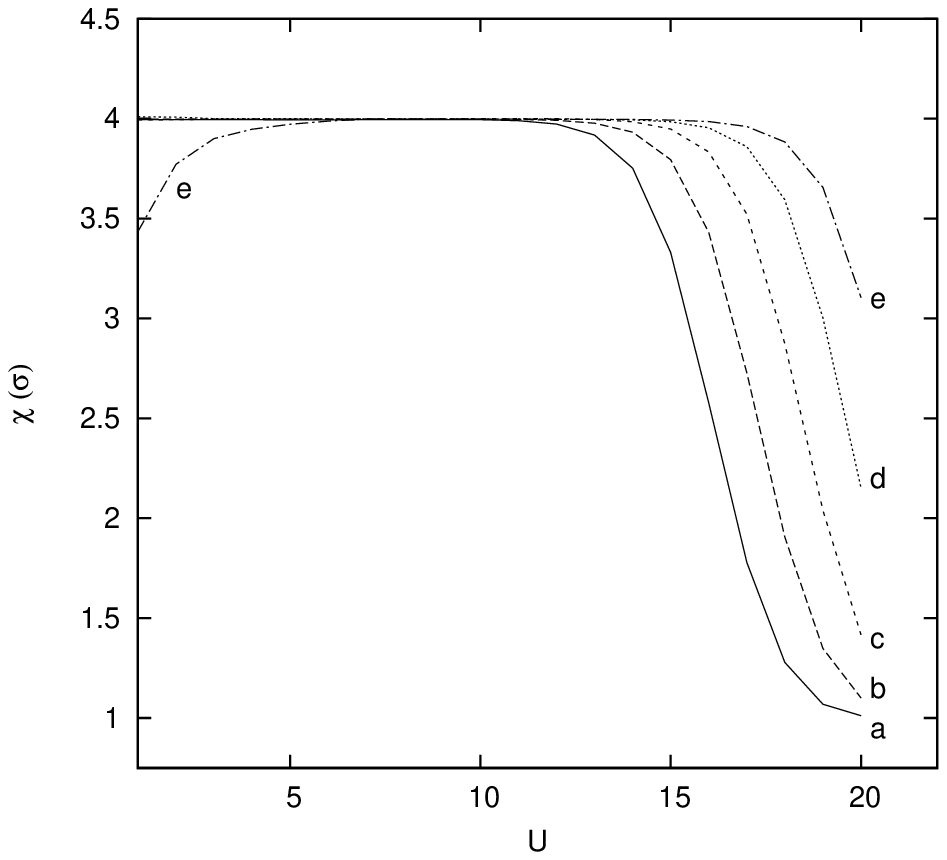}
\caption{Convergence factors for $\chi (\sigma )$ (see text for
definition) along lines of constant $u$ for non-AMR simulations
with varying resolutions. Lines are for resolutions a) $N=5$, b)
$N=10$, c) $N=20$, d) $N=40$ and e) $N=80$. \label{fig:B1}}
\end{figure}
\begin{figure*}
\subfigure[$||C_{vv}||$ (see text) for the constraint equation
\eqref{eq:7}.
\label{fig:B2a}]{\includegraphics[width=0.49\textwidth]{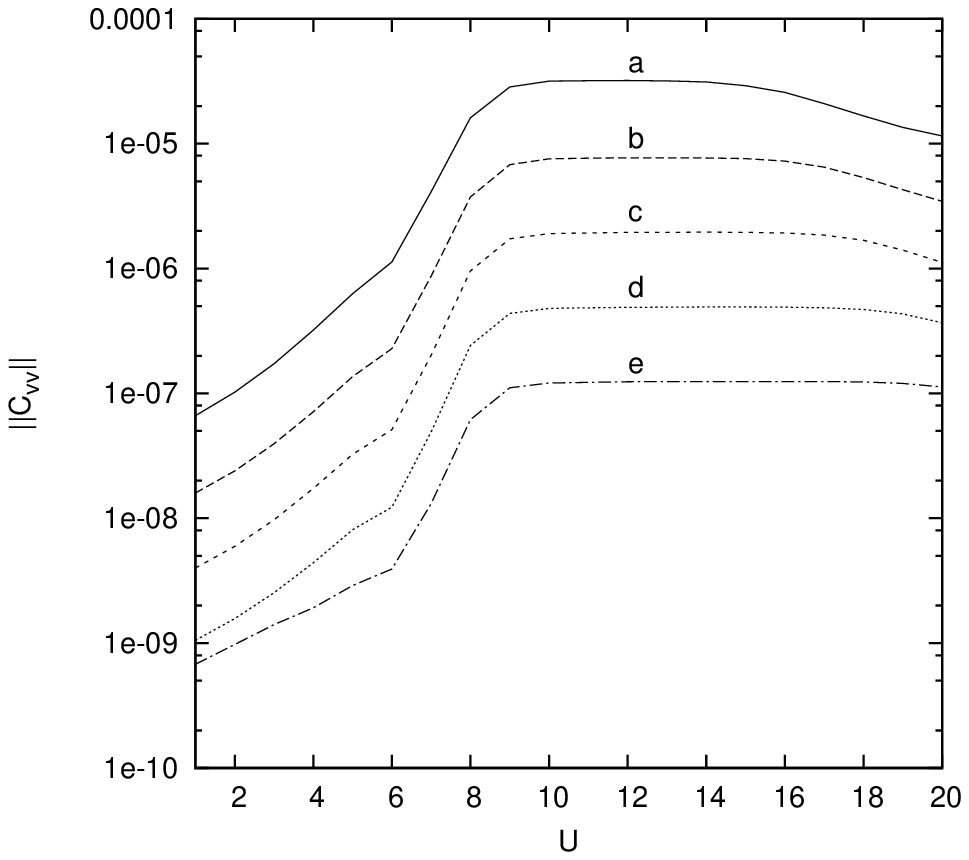}}
\subfigure[$||C_{uu}||$ (see text) for the constraint equation
\eqref{eq:8}.
 \label{fig:B2b}]{\includegraphics[width=0.49\textwidth]{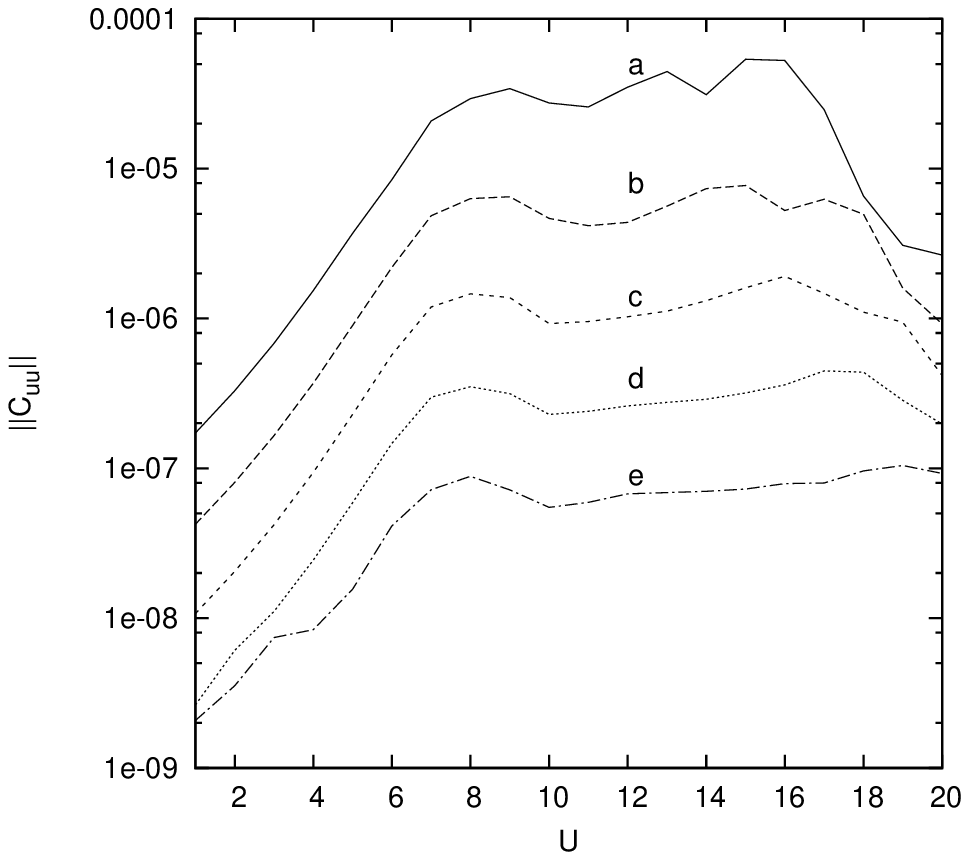}}
\caption{{\label{fig:B2} Average values of the constraint
equations along lines of constant $v$ for non-AMR simulations with
varying resolutions. Lines correspond to resolutions of a) $N=5$,
b) $N=10$, c) $N=20$, d) $N=40$ and e) $N=80$.}}
\end{figure*}

First we test the basic convergence properties of the code by
performing a series of simulations with varying base resolutions
but with the AMR algorithm disabled.

To estimate the convergence rate along a ray of $u=constant$ we
calculate :

\begin{equation}
  \label{eq:chidef}
  \chi (x_N) \equiv \frac{1}{n} \frac{\sum_i^n |x_N^i - x_{2N}^i|}{\sum_i^n
|x_{2N}^i - x_{4N}^i|}
\end{equation}

where $x_N^i$ denotes the dynamic variable $x$ at the $i$-th grid
point (along the given $u=constant$ line) of simulation with base
resolution $N$ (where $N\equiv 1/\Delta u =1/\Delta u$). The $n$
points over which we sum, is taken along an outgoing ray of
$u=constant$. Naturally, in order for this expression to make
sense, we are only using those $n$ points which all three
resolutions have in common. If the code is second-order
convergent, we would expect to see a convergence factor of $\chi =
4$.

\begin{figure*}
\subfigure[Convergence factor $\xi (r)$.
\label{fig:B3a}]{\includegraphics[width=0.49\textwidth]{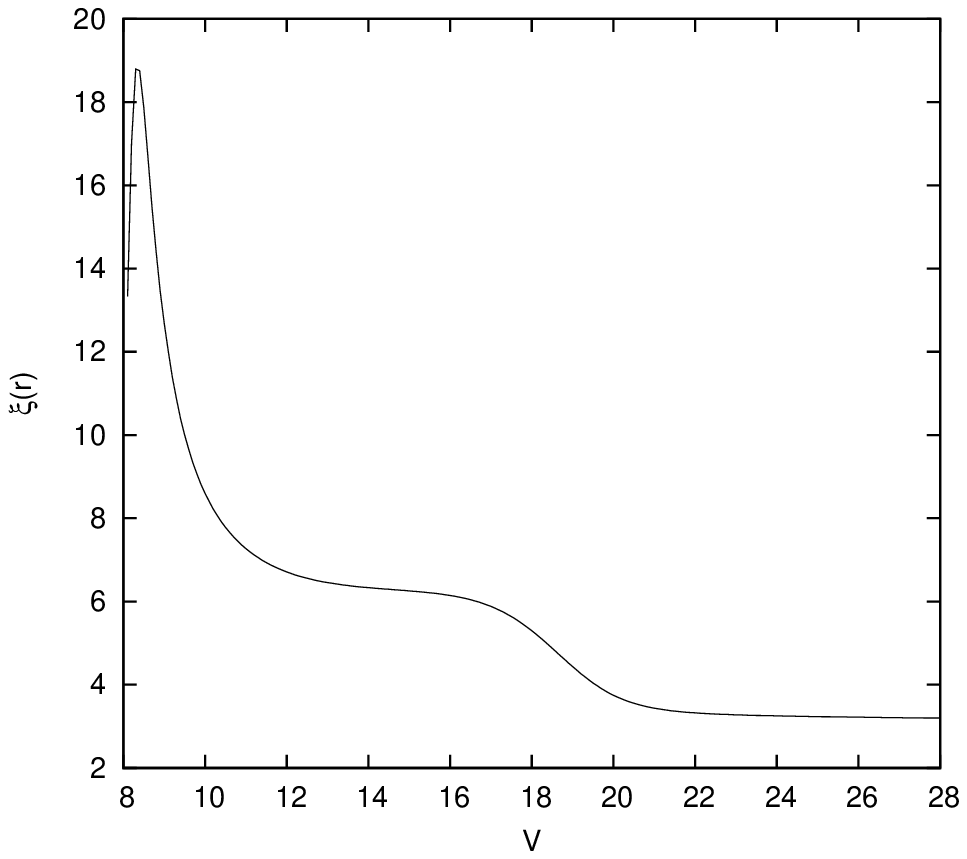}}
\subfigure[Convergence factor for $\xi (\sigma)$.
 \label{fig:B3b}]{\includegraphics[width=0.49\textwidth]{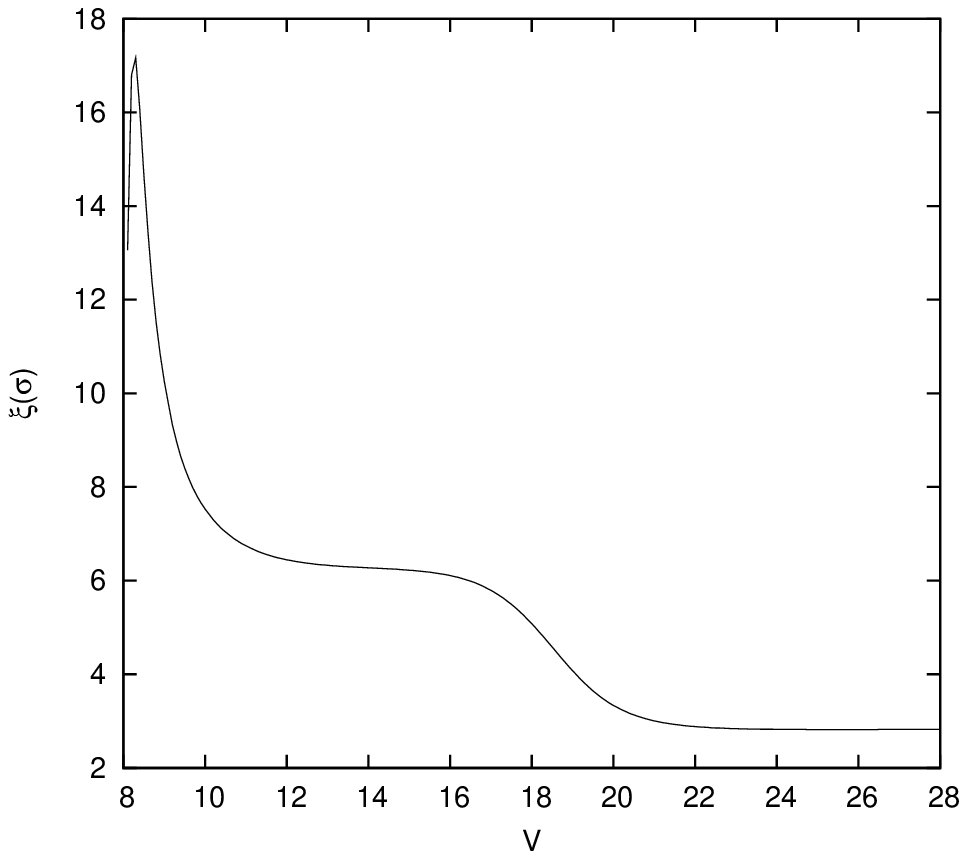}}
\subfigure[Convergence factor for $\xi (\phi)$.
\label{fig:B3c}]{\includegraphics[width=0.49\textwidth]{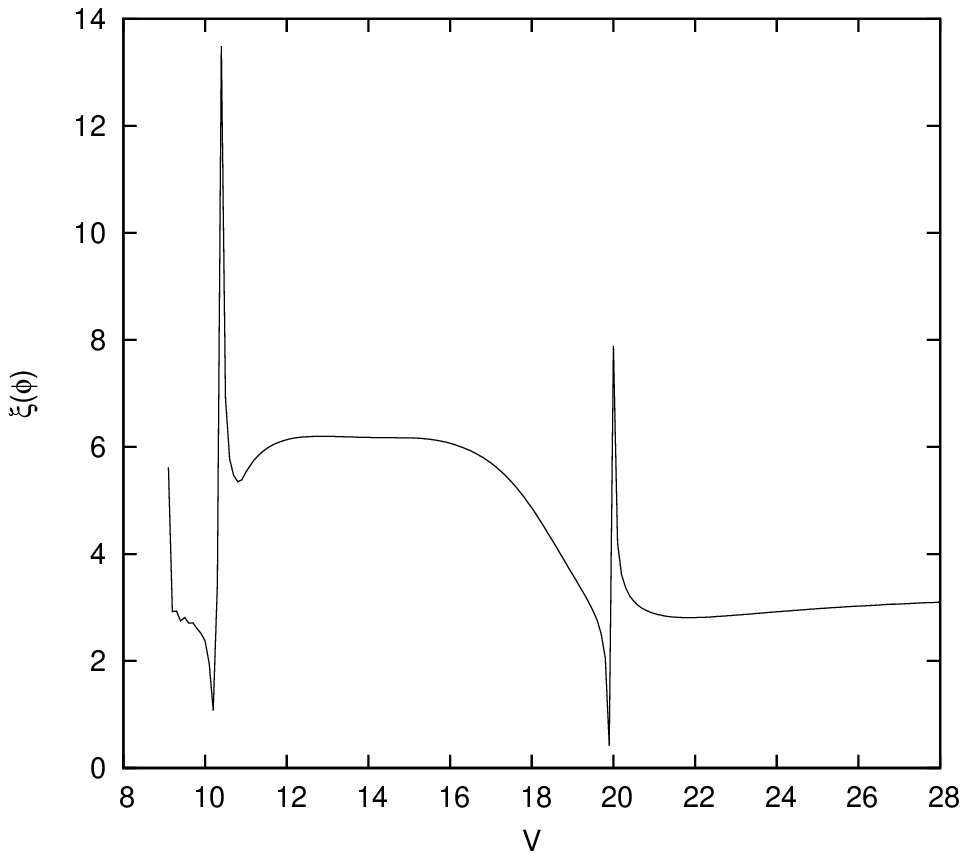}}
\subfigure[Convergence factor for $\xi (\psi)$...
 \label{fig:B3d}]{\includegraphics[width=0.49\textwidth]{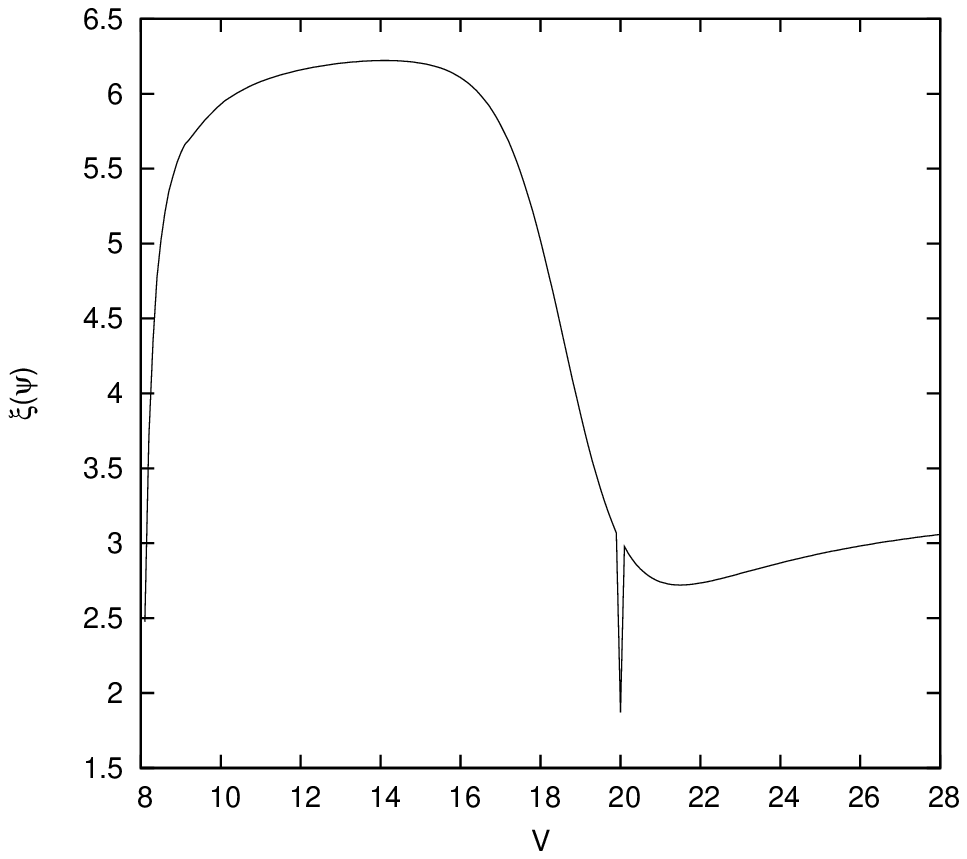}}
\caption{{\label{fig:B3} Convergence factors $\xi (x)$ along line
of $u=20$ for simulations with AMR enabled, see text for
details.}}
\end{figure*}

We calculate expression \eqref{eq:chidef} along lines of many
$u=constant$ and for many base resolutions $N$, such that we get a
picture of the convergence properties throughout all of our
computational domain for varying resolutions. Figure \ref{fig:B1}
shows the convergence rates for the dynamical variable $\sigma$
for a large number of resolutions.  The figure shows that for
small $u$ (below $u=10$), the code is second order converging. The
line $e$ (corresponding to the base resolution $N=80$) gives the
impression of less than second order convergence at small $u$,
however a closer investigation of this behavior has revealed that
this is due to the solution in this region has reached such a high
degree of convergence, that numerical roundoff errors in the
numerical representation affects the calculation of the
convergence factor.

For high $u$ we see that especially for low resolutions, the
convergence is far from second order. This is mainly due to the
solution in this region, for low resolutions, is diverging too
much from the physical solution. Such behavior is not uncommon for
non-linear systems. However, as is seen, for increasing
resolutions, even for high $u$, the convergence factor of the
solution becomes increasingly higher, indicating that the
numerical solution is converging with second order accuracy. That
a fixed resolution can be more than adequate in one part of the
computational domain and insufficient in another part is a strong
indication of the need for AMR in our code. The properties of the
code with AMR enables is investigated in the next subsection.

The other three dynamical variables has convergence properties
similar to those of $\sigma$.

It is not enough, however, to demonstrate that our code is
converging, we also need to demonstrate that it is converging to a
physical solution. We do this by calculating the average value
\begin{equation}
  ||x_N|| = \frac{1}{n}\sum_i^n x_N^i
\end{equation}

of the two constraint equations along lines of constant $u$ for
different resolutions. Figure \ref{fig:B2} demonstrates that both
of the constraint equations is converging towards zero for
increasing resolution, thus demonstrating that our numerical
solution is indeed converging towards a physical solution for
increased resolutions.

\subsection{AMR convergence tests}

An important component of our code is its AMR capabilities. In
this subsection we demonstrate the behavior of the code when the
AMR algorithm is enabled.

For each simulation we must specify a resolution of the underlying
base grid and we must also specify a truncation error acceptance
limit ($TE_{max}$) which determines whether or not a computation
cell should be split. For all simulations we allow for a maximum
of 8 splitting levels (including the basegrid and it's
self-shadowing level, see Appendix \ref{sec:appA} and \cite{Pretorius04}).

\begin{figure*}
\subfigure[Average of $C_{vv}$ along lines of constant $u$ for
varying $TE_{max}$..
\label{fig:B5a}]{\includegraphics[width=0.49\textwidth]{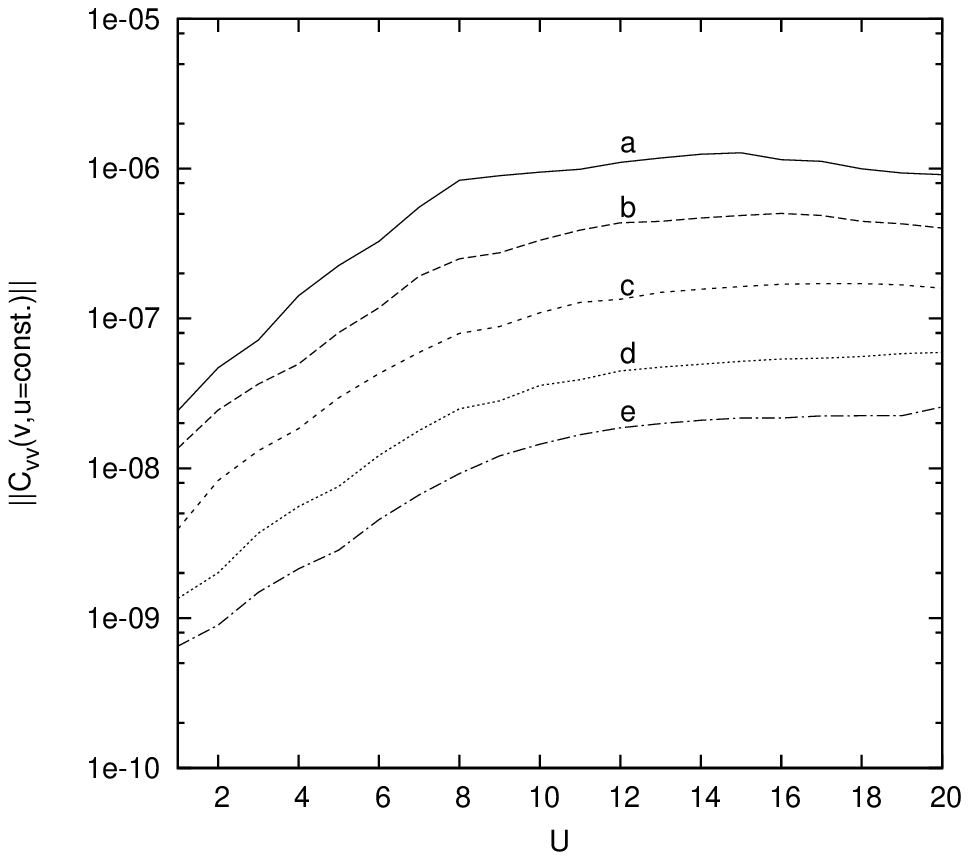}}
\subfigure[Average of $C_{uu}$ along lines of constant $u$ for
varying $TE_{max}$.
 \label{fig:B5b}]{\includegraphics[width=0.49\textwidth]{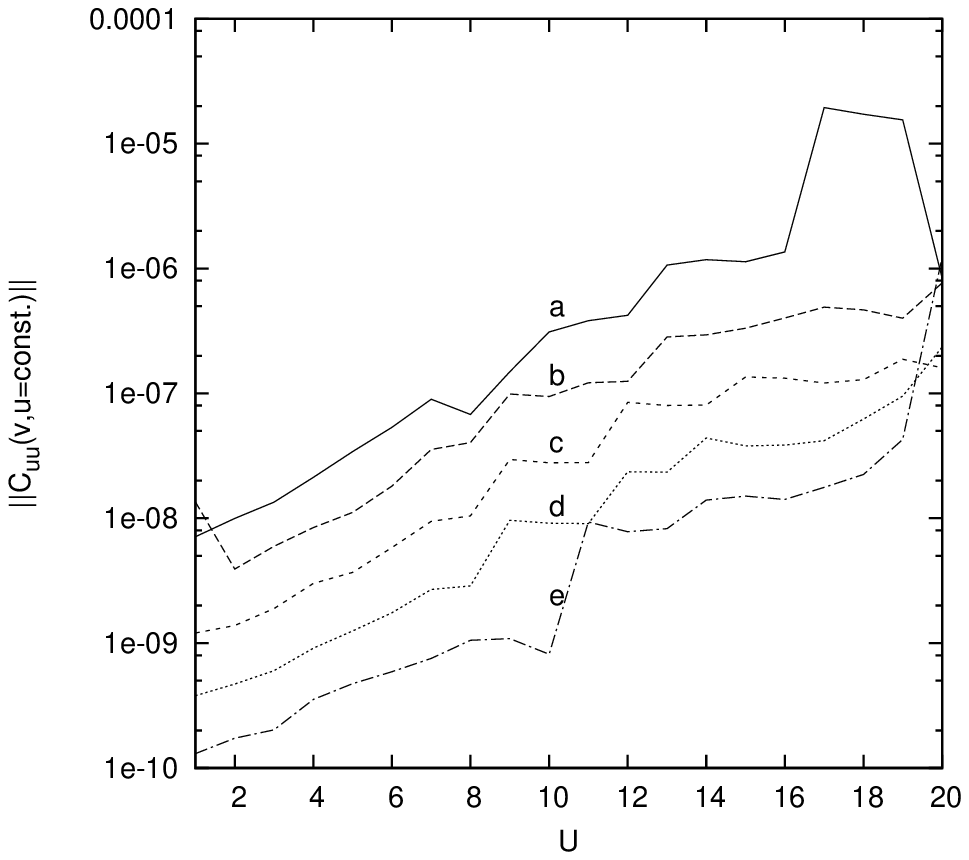}}
\caption{{\label{fig:B5} Average residual of the constraint
equations (along lines of constant $u$) for fixed base resolution
$N=5$, AMR enabled and varying $TE_{max}$. Lines are for a) $TE_{max} = 1e-8$,
b) $TE_{max} = 1e-9$, c) $TE_{max} = 1e-10$, d) $TE_{max} = 1e-11$
and e) $TE_{max} = 1e-12$.}}
\end{figure*}

To investigate the convergence properties for the code with AMR
enabled we did three simulations with the same initial conditions
as in the previous section, varying the base resolution and $TE_{max}$ to mimic the doubling of resolution from one
simulation to the next as follows; The lowest resolution
simulation has a base resolution of $N=5$ (corresponding to a
total of 100 gridpoints) and a $TE_{max}$ of $\tau_N$, the second
simulation has a base resolution of $N=10$ and a $TE_{masx}$ of
$\tau_{2N} = \tau_N / 4$ and the last simulation has a base
resolution of $N=20$ and a $TE_{masx}$ of $\tau_{4N} = \tau_N / 16$.
The varying of $TE_{max}$ mimics the doubling of the resolution
from one simulation to the next (under the assumption of second
order convergence) and in principle the splitting level hierarchies
should be similar for all three simulations. In practice, however,
we do not obtain identical splitting hierarchies because the
splitting structure is determined by the TE \textit{estimate} and
this will not scale exactly as the leading order part of the
actual truncation error, which decreases by a factor of 4 each
time the mesh spacing is halved for a second order accurate scheme
\cite{Pretorius04}. Nevertheless, this test gives a good
indication of the behavior of the code and if the test showed
non-convergent results it would indicate problems with the
implementation.

In this section, to calculate the convergence factor, we use data
along $u=20$ (i.e. the upper border of our computational domain).
As this is the outer border of our computational domain,
convergence here is a good indication of the behavior throughout
our domain. Furthermore, this border is closest to the $r=0$
singularity (that would be formed within our domain if the initial
$\Phi$-pulse had been slightly stronger), hence it is the most
demanding place in our domain to test convergence.

Figure \ref{fig:B3} shows the convergence factors :
\begin{equation}
  \label{eq:xidef}
  \xi (x_N) \equiv \frac{1}{n} \frac{|x_N^i - x_{2N}^i|}{ |x_{2N}^i -
x_{4N}^i|}
\end{equation}
where $x_N^i$ denotes the dynamic variable $x$ at the $i$-th grid
point (along $u=20$) of simulation with base resolution $N$ (where
$N\equiv 1/\Delta u =1/\Delta u$). Naturally, in order for this
expression to make sense, we are only using those $n$ points which
all three resolutions have in common. If the code is second-order
convergent, we would expect to see a convergence factor of $\xi =
4$.

In the figure, we see that for small $v$, in general, the factors
are greater than $4$, whereas for larger $v$, in general the
convergence factors are smaller. It is clear that, especially for
$r$ and $\sigma$ we are seeing unrealistically high convergence
factors for early $v$, however, as noted above, it would be
surprising to recover perfect second order convergence in this
test. The important point of fig. \ref{fig:B3}, however, is that
all the dynamic variables are converging, also when the AMR
algorithm is turned on, and that this convergence is in good
agreement with that expected.

Finally we show the results of a series of simulations with a
fixed base resolution of $N = 5$ but varying the $TE_{max}$
limits. In fig. \ref{fig:B5} is shown the average of each of the
constraint equations along lines of constant $u$, i.e. :

\begin{equation}
  \label{eq:figB5def}
  ||x_{TE}|| = \frac{1}{n} \sum_i^n |x_{TE}^i |
\end{equation}
(where $x$ is the constraint equation, $i$ denotes the $i$-th out
of $n$ point along a line of constant $u$).

We see that both the constraint equations are converging towards
zero (although, for this test is makes no sense to talk about the
"convergence rate"). It is also visible from fig. \ref{fig:B5},
that for increasing $u$, the constraint equations are increasingly
violated. This is only expected and is caused partly by truncation
errors being accumulated throughout the computational domain and
partly by the solution getting close to the region where a black
hole is formed and hence also the region where strong gradients in
the dynamic variables puts larger requirements on the numerical
code.

\begin{figure}
\includegraphics[width=0.49\textwidth]{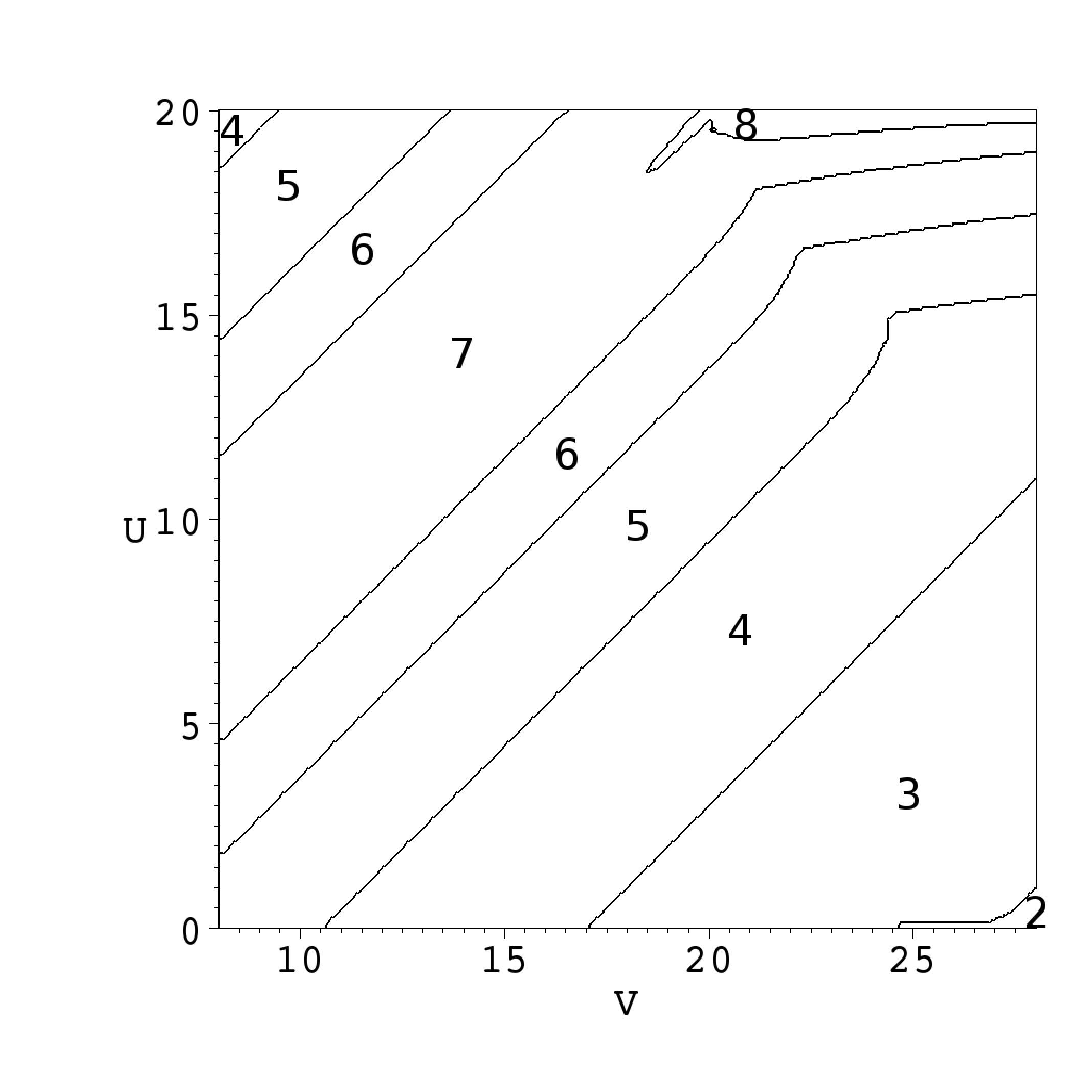}
\caption{Demonstration of splitting level structure. Numbers
indicate the number of split levels including the base shadow hierarchy (i.e. ``2'' is most basic structure possible in our configuration).\label{fig:B4}}
\end{figure}

Another important point related with an AMR algorithm is how
smooth the splitting levels are. It is well known that when a
higher splitting level is introduced, this usually introduce high
frequency noise into the solution near the borders of higher
resolution \cite{Pretorius04}, which, in the worst case, may
ultimately lead to the crash of the code. Because of this, it is
highly desirable to have as few splitting levels as possible and
to have as few splitting borders as possible. Figure
\ref{fig:B4} illustrates the splitting level structure for a
typical simulation with base resolution and $TE_{max}$ equal to
the settings used in our actual simulations in sections
\ref{sec:IV}-\ref{sec:VI}.

It is seen that the splitting structures are very smooth with no
sudden or sporadic jumps, indicating that our implementation of
the AMR algorithm is very successful. In \cite{Pretorius04} it is
suggested to use a small numerical dissipation, but we have found
that this is not necessary with our code, for the simulations
presented in this paper.

Finally it should be noted, that for all results presented in this
paper, we have performed a great number of simulations with
varying base resolutions and $TE_{max}$. This we did to be
certain that the results on which we have based our conclusions,
had converged to such a degree that we could trust the results to
be representative of the underlying physics.

\bibliography{paper}

\end{document}